\documentclass[twocolumn,tighten,trackchanges,twocolappendix]{aastex701}
\usepackage{CJK}
\usepackage{amsmath}
\usepackage{multirow}

\usepackage{xspace}

\newcommand{\Ni}{$^{56}$Ni\xspace}

\defcitealias{millerZTFEarlyObservations2020}{M20}

\graphicspath{{./}{}}

\begin{document}
\title{Decoding the Early-Time Light Curves of Type Ia Supernovae. I. A Hierarchical Bayesian Framework for Demographic Inference}

\author[0000-0002-7866-4531,gname=Chang,sname=Liu]{Chang~Liu \begin{CJK*}{UTF8}{gbsn}(刘畅)\end{CJK*}}
\affil{Department of Physics and Astronomy, Northwestern University, 2145 Sheridan Rd, Evanston, IL 60208, USA}
\affil{Center for Interdisciplinary Exploration and Research in Astrophysics (CIERA), Northwestern University, 1800 Sherman Ave, Evanston, IL 60201, USA}
\affil{NSF-Simons AI Institute for the Sky (SkAI), 172 E. Chestnut St., Chicago, IL 60611, USA}
\email{ptg.cliu@u.northwestern.edu}

\author[0000-0001-9515-478X]{Adam~A.~Miller}
\affil{Department of Physics and Astronomy, Northwestern University, 2145 Sheridan Rd, Evanston, IL 60208, USA}
\affil{Center for Interdisciplinary Exploration and Research in Astrophysics (CIERA), Northwestern University, 1800 Sherman Ave, Evanston, IL 60201, USA}
\affil{NSF-Simons AI Institute for the Sky (SkAI), 172 E. Chestnut St., Chicago, IL 60611, USA}
\email{amiller@northwestern.edu}


\correspondingauthor{Chang~Liu}
\email{ptg.cliu@u.northwestern.edu}

\begin{abstract}
Light curves of Type Ia Supernovae (SNe\,Ia) in the days following explosion encode the diversity of progenitor systems and explosion physics.
We present a hierarchical Bayesian framework to robustly constrain the population-level light-curve morphology of SNe\,Ia by fitting a large light-curve dataset simultaneously to power-law rises.
Using a multivariate Gaussian population prior, this framework automatically down-weights sparsely sampled SNe and noisy measurements in the inference, obviating the need for restrictive quality cuts that introduce selection biases.
Validation on simulated power-law light curves demonstrates that the population prior effectively suppresses the volume-projection bias from the asymmetric likelihood:
compared to the classic two-step approach of fitting individual SNe and then aggregating the results, the hierarchical approach dramatically reduces the bias on the population-level parameters (mean, scatter, and correlation).
When fitting the power-law model to light curves with more realistic morphologies, while the rise time can be mildly underestimated due to model misspecification, the recovered population scatter remains reliable. Furthermore, SNe with early flux excesses can emerge as outliers in the inferred parameter space, offering a potential diagnostic for identifying such events.
Finally, we show that the inferred population distribution can also improve individual-event inference. Restricting the population prior to nuisance amplitudes, while preserving the complete correlation structure, regularizes fits to individual SNe without shrinking the physically meaningful rise time and rise index toward their population means.
\end{abstract}



\section{Introduction} \label{sec:intro}

The wealth of data from modern time-domain surveys has shifted the study of Type Ia Supernovae (SNe\,Ia) from observing individual events to statistically characterizing large populations. While their remarkably standardizable light curves near maximum luminosity have enabled the triumph of precision cosmology \citep{riessObservationalEvidenceSupernovae1998,perlmutterMeasurements42HighRedshift1999,abbottDarkEnergySurvey2024}, this standardization obscures physical insights into diverse progenitor channels of SNe\,Ia as explosions of carbon-oxygen white dwarfs (WDs) in binary systems.

Recent efforts to understand the explosion physics of SNe\,Ia have increasingly focused on the rich diversity and subtle parameter correlations using observations far from peak brightness. In particular, data collected within the first few days after explosion offer a unique window into the outermost ejecta and the immediate circumstellar environment. 
Intriguingly, at these early phases, SNe\,Ia exhibit notably more diversity than near maximum light, displaying a wide range of rise times \citep{firthRisingLightCurves2015,zhengEmpiricalFittingMethod2017a, millerZTFEarlyObservations2020} and color evolution trends \citep{stritzingerRedBlueEarly2018,bullaZTFEarlyObservations2020,niInfantTypeIa2025}, which trace underlying variations in kinetic energy, opacity, and \Ni distribution in the outer ejecta \citep{piroWHATCANWE2013,piroCONSTRAINTSSHALLOW562014}. Furthermore, early-time data have unveiled a variety of flux excesses departing from a smooth rise \citep[e.g.,][]{zhengVERYYOUNGTYPE2013, marionSN2012cgEVIDENCE2016,hosseinzadehEarlyBlueExcess2017, millerEarlyObservationsType2018,dimitriadisK2ObservationsSN2018,liPhotometricSpectroscopicProperties2018,ashallSpeedBumpSN2022,saiObservationsVeryYoung2022,wangFlightBumblebeeEarly2024,iskandarSN2021hprNormal2025}. These anomalies can probe interactions between the expanding ejecta and a non-degenerate companion \citep{kasenSEEINGCOLLISIONSUPERNOVA2009} and/or circumstellar material \citep[CSM;][]{piroEXPLORINGPOTENTIALDIVERSITY2016}, constraining the configuration and mass-transfer history of the binary system.

However, directly inferring physical parameters from early-time light curves is hindered by the scarcity of explosion models suitable for rigorous posterior sampling \citep[see][for a pilot study]{mageeDetermining56Ni2020}. To circumvent this limitation, early-time observables are typically distilled into key metrics by fitting the rising light curves with simple parametric models. These empirical representations, such as the rise time, rising slope, and color evolution, can then be leveraged to characterize population demographics and uncover underlying morphological correlations. 

The most widely adopted parametric form is a single power-law rise:
\begin{equation} \label{eq:powerlaw}
f(t)\propto (t-t_\mathrm{fl})^\alpha, 
\end{equation}
where $t_\mathrm{fl}$ is the time of first light\footnote{Note that $t_\mathrm{fl}$ is distinct from the true explosion time due to a ``dark phase'' during which photons from radioactive decay diffuse through the ejecta before escaping the photosphere \citep{piroWHATCANWE2013,piroCONSTRAINTSSHALLOW562014}.} and $\alpha$ is the power-law rise index dictating the rate of flux growth.
This model is motivated by the expanding fireball picture, where the broad-band optical emission in the Rayleigh-Jeans tail can be well approximated by $f\propto R^2T$, with $R$ and $T$ being the photospheric radius and temperature. Assuming a constant photospheric velocity and temperature at early phases leads to a quadratic rise \citep[$\alpha=2$;][]{riessRiseTimeNearby1999}. However, these assumptions typically break down within the first few days after explosion: the photospheric velocity may decrease rapidly \citep{maguireExploringSpectralDiversity2014}, and temperature evolution varies remarkably \citep{stritzingerRedBlueEarly2018,bullaZTFEarlyObservations2020,niInfantTypeIa2025}. In fact, synthetic light curves from realistic explosion models are usually not well described by a single power-law \citep[e.g.,][]{noebauerEarlyLightCurves2017}.
Nevertheless, early studies fitting stretch-corrected light curves to a common fiducial template generally found a typical rise index $\alpha\simeq2$, and a typical rise time $t_\mathrm{rise}$ in the range $\simeq16$--19\,days \citep{alderingRiseTimesHigh2000,goldhaberTimescaleStretchParameterization2001,conleyRiseTimeType2006,ganeshalingamRisetimeDistributionNearby2011,gonzalez-gaitanRISETIMENORMAL2011}.

With a growing number of well-sampled early light curves, fitting individual SNe\,Ia without assuming a universal $\alpha$ or perfect rise-decline correlation has revealed pronounced diversity in rise indices \citep{firthRisingLightCurves2015,zhengEmpiricalFittingMethod2017a,papadogiannakisRbandLightcurveProperties2019,millerZTFEarlyObservations2020,fausnaughFourYearsType2023}. This variance traces underlying differences in kinetic energy, opacity, and \Ni distribution within the outermost ejecta \citep{piroWHATCANWE2013,piroCONSTRAINTSSHALLOW562014}. Furthermore, although rise and decline timescales loosely correlate, the relationship is imperfect, indicating that a single stretch parameter cannot fully capture early-time behavior \citep{haydenRISEFALLTYPE2010,gonzalez-gaitanRISETIMENORMAL2011}.

The advent of wide-field time-domain surveys, such as the All-Sky Automated Survey for Supernovae \citep[ASAS-SN;][]{shappeeMANCURTAINXRAYS2014}, the Asteroid Terrestrial-impact Last Alert System \citep[ATLAS;][]{tonryATLASHighcadenceAllsky2018}, and the Zwicky Transient Facility \citep[ZTF;][]{bellmZwickyTransientFacility2018}, has revolutionized this landscape by routinely capturing multi-band observations within the first week of SN explosions. 
Despite these observational advances, extracting robust morphological parameters remains a fundamental challenge. Early-time observations are often sparse, heteroscedastic, and subject to complex selection effects. Consequently, attempting to characterize vast, heterogeneous populations by fitting noisy light curves in isolation frequently yields overconfident or severely biased results. For instance, \citeauthor{millerZTFEarlyObservations2020} (\citeyear{millerZTFEarlyObservations2020}; hereafter \citetalias{millerZTFEarlyObservations2020}) demonstrated that modeling the early-time light curves of individual ZTF SNe independently produces an unphysical correlation between the inferred rise time and redshift driven by the Malmquist bias. Although limiting the analysis to a ``golden'' subset of well-sampled SNe can improve the reliability of individual fits \citep[e.g., \citetalias{millerZTFEarlyObservations2020},][]{fausnaughFourYearsType2023}, these stringent quality cuts inherently inject new selection biases into the population demographics, which are rarely quantified.

Hierarchical Bayesian modeling offers a principled solution to these challenges. Under this framework, each celestial event is treated as a draw from an underlying population distribution that is simultaneously inferred from the full dataset, automatically assigning greater weight to well-constrained events while still leveraging information from noisy observations.
A key feature known as ``Bayesian shrinkage'' allows the model to borrow strength across the ensemble: the posterior distributions of poorly constrained events are shrunk toward the population mean, yielding more precise individual constraints and reducing the overall mean squared error. 
This methodology has a distinguished record across astrophysics---from the mass and spin distributions of compact-object binaries in gravitational wave astronomy \citep{thraneIntroductionBayesianInference2019,mandelExtractingDistributionParameters2019} to exoplanet demographics \citep{hoggINFERRINGECCENTRICITYDISTRIBUTION2010,foreman-mackeyEXOPLANETPOPULATIONINFERENCE2014}, SN\,Ia standardization and cosmology \citep{mandelTYPEIaSUPERNOVA2011, hintonSteveHierarchicalBayesian2019}, and weak gravitational lensing \citep{thedesandsptcollaborationsSPTClustersHST2024}.

In this paper, we present a hierarchical Bayesian framework to robustly infer population-level properties from early-time SN\,Ia light curves. This approach enables the direct modeling of population statistics (specifically the mean, scatter, and correlation) for rise times, rise indices, and early color evolution. As we demonstrate, this method yields significantly less biased inferences compared to classical approaches that simply aggregate the fits of individual events. 

As the first paper in a series exploring early-time SN\,Ia demographics, we focus here on the technical foundation and validation of our model. In Section~\ref{sec:setup}, we revisit the parametric power-law rise and detail the mathematical formulation and implementation of our hierarchical Bayesian framework. In Section~\ref{sec:mock_pwl}, we validate the model's ability to recover true population statistics using extensive datasets of synthetic power-law rises, while in Section~\ref{sec:mock_misspecification} we explore the impact of fitting a power-law model to light curves with more realistic, complicated morphologies. Although the hierarchical Bayesian framework is designed to infer the population-level properties, Section~\ref{sec:ind} explores the potential of using the inferred population distribution as a prior to improve the constraints on individual SNe. 
We summarize our findings in Section~\ref{sec:conclusions}.

\section{Setting the Stage} \label{sec:setup}
Following \citetalias{millerZTFEarlyObservations2020}, we characterize the early-time emission in filter $b$ from SN $k$ as
\begin{equation}
    f_{k, b}(t) = C_{k, b} + H(t-t_{\mathrm{fl}, k})A_{k,b}\left(\frac{t-t_{\mathrm{fl}, k}}{t_\mathrm{pivot}}\right)^{\alpha_{k,b}},
\end{equation}
where $f_{k, b}(t)$ is the observed single-band flux at time $t$ normalized by the maximum flux of SN $k$ measured in the same filter, $t_{\mathrm{fl},k}$ is the time of first light, $A_b$ is a normalization factor in filter $b$ which we refer to as the amplitude, $\alpha_b$ is the power-law rise index in filter $b$ characterizing the rise index,\footnote{Throughout the paper, we assume $\alpha>1$, such that its derivative is well-defined at $t_\mathrm{fl}$ to avoid stiffness in the posterior sampling.
This is a reasonable assumption for normal SNe\,Ia whose early rises are generally smooth and convex.} $C_{k, b}$ is a constant representing the baseline flux present in the reference image prior to the SN discovery,
and $H$ is the Heaviside step function ensuring $f = 0$ before $t_\mathrm{fl}$. 

We fit the early-time light curves in both $g$ and $r$ filters up to the epoch when the flux reaches a certain fraction of the peak flux, which we refer to as the truncation threshold. \citetalias{millerZTFEarlyObservations2020} demonstrated that below the truncation threshold of 40\%, the early-time SN\,Ia light curves can be reasonably well fit by a power-law, although the choice of this threshold can impact the inferred population-level parameters (see Section~\ref{sec:mock_frac}).
To improve the sampling efficiency, we introduce a constant pivot time $t_\mathrm{pivot}$ which reduces the covariance between $A_b$ and $\alpha_b$, which is implicitly defined to be 10\,days in \citetalias{millerZTFEarlyObservations2020}. Since $f_{k,b}$ is normalized by the peak flux, setting $t_\mathrm{pivot}$ to be the typical rise time to the truncation threshold ensures minimal covariance between these two parameters. We find that the rise time to 40\% of the peak flux is $\sim$8\,days in the ZTF SN\,Ia sample, and therefore fix $t_\mathrm{pivot}=8$\,days in this work.

Also following \citetalias{millerZTFEarlyObservations2020}, we assume that $t_{\mathrm{fl}, k}$ is the same across all filters for each SN\,Ia $k$, while $A_{k,b}$, $\alpha_{k,b}$, and $C_{k,b}$ are both SN- and filter-specific.

\subsection{Hierarchical Bayesian Modeling} \label{sec:hbm}

\begin{figure*}[ht!]
    \centering
    \includegraphics[width=0.95\linewidth]{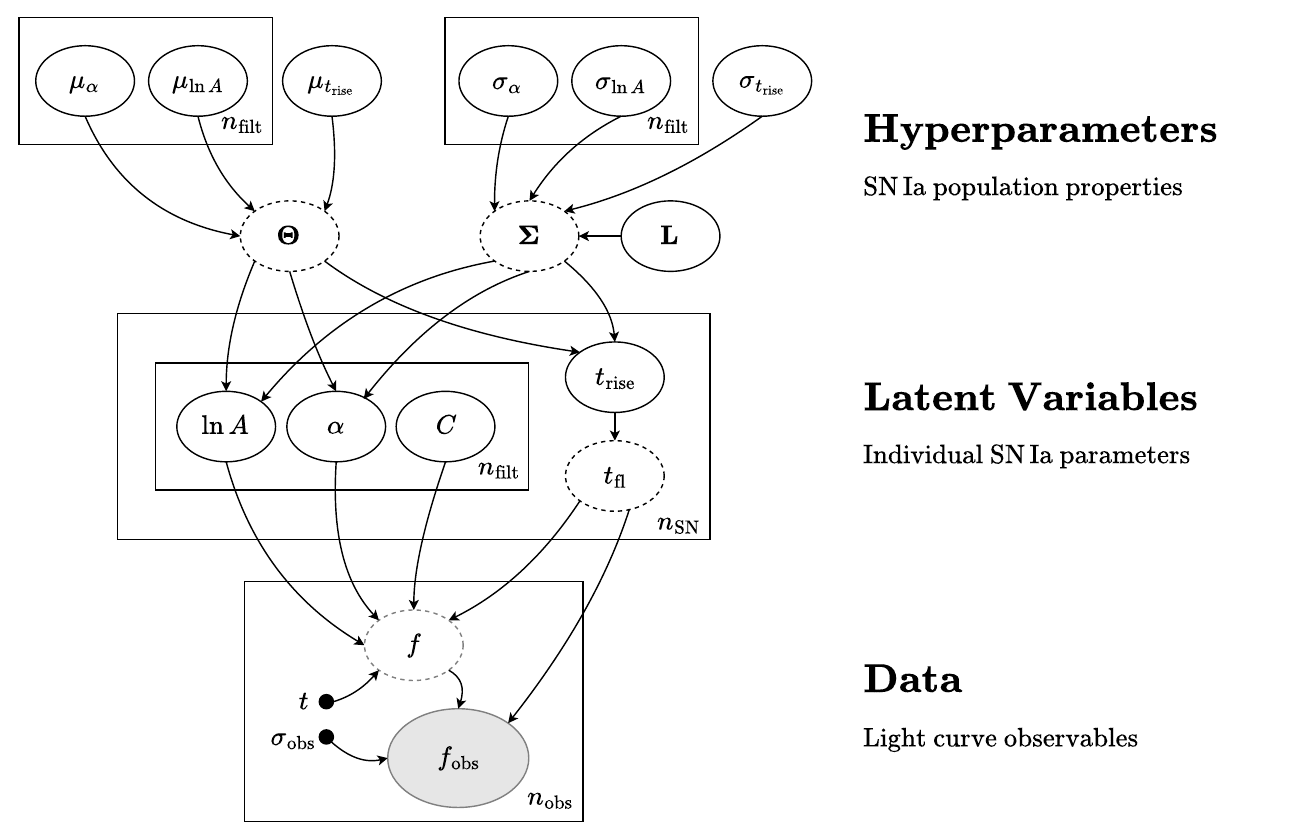}
    \caption{Probabilistic graphical model of our hierarchical Bayesian framework for modeling early-time SN\,Ia light curves. Nodes (ellipses) represent variables: solid ellipses are random variables sampled from the priors listed in Table~\ref{tab:priors}, dashed ellipses are deterministic quantities derived from parent nodes; black dots represent fixed constants measured externally; and the shaded ellipse indicates the observed flux, $f_\mathrm{obs}$. Plates (boxes) indicate replication over SNe ($n_\mathrm{SN}$), filters ($n_\mathrm{filt}$), and observations ($n_\mathrm{obs}$). Arrows show conditional dependencies.}
    \label{fig:hierarchical_model}
\end{figure*}

To overcome the biases introduced by aggregating fits to individual SNe, we develop a hierarchical Bayesian model to simultaneously model all SNe\,Ia in a volume-complete, unbiased sample. In this framework, we assume that the light-curve parameters reflecting the physical properties of the outermost SN ejecta are drawn from a common joint distribution. This approach allows us to leverage the information from the entire sample to constrain the population-level parameter distributions, while also accounting for the uncertainties in individual measurements.

Specifically for each SN\,Ia $k$, the rise time $t_\mathrm{rise, k} = t_{0, k} - t_{\mathrm{fl}, k}$, where $t_{0, k}$ is the epoch of the rest-frame $B$-band maximum as inferred by \texttt{SALT2} \citep{guySALT2UsingDistant2007}, the power-law rise index $\alpha_{k, b}$, and the logarithmic amplitude $\ln A_{k, b}$ in each filter $b$ ($=g, r$) are drawn from a multivariate Gaussian distribution:
\begin{equation}
    \boldsymbol{\theta}_k \sim \mathcal{N}(\boldsymbol{\Theta}, \boldsymbol{\Sigma}),
\end{equation}
where for SN $k$,
\begin{equation}
    \boldsymbol{\theta}_k = (t_{\mathrm{rise}, k}, \alpha_{g, k}, \alpha_{r, k}, \ln A_{g, k}, \ln A_{r, k})^\mathsf{T},
\end{equation}
and $\boldsymbol{\Theta}$ is the population mean vector defined as:
\begin{equation}
    \boldsymbol{\Theta} = (\mu_{t_{\mathrm{rise}}}, \mu_{\alpha_g}, \mu_{\alpha_r}, \mu_{\ln A_g}, \mu_{\ln A_r})^\mathsf{T}.
\end{equation}
All these components are drawn from uniform priors over physically reasonable ranges. The covariance matrix $\mathbf\Sigma$ is drawn from an LKJ prior \citep{lewandowskiGeneratingRandomCorrelation2009} as:
\begin{equation}\label{eq:Sigma}
\boldsymbol{\Sigma}
= \mathbf{L}\,
\mathrm{diag}(\boldsymbol{\sigma}^2)\,
\mathbf{L}^\mathsf{T},
\end{equation}
where $\mathbf{L}$ is the Cholesky decomposition of an underlying correlation matrix with an LKJ prior, a probability distribution over valid correlation matrices that allows regularizing the expected degree of correlation among parameters.
We set the LKJ shape parameter to $\eta=1$, corresponding to a uniform prior on the correlation matrix. With this setup, the prior on the correlation coefficient $\rho$ between any pair of $\boldsymbol{\theta}_k$ components is
\begin{equation}
P(\rho) \propto {(1-\rho^2)^{\eta - 1 + (n_\mathrm{dim}-2)/2}}.
\end{equation}
With the number of dimensions $n_\mathrm{dim}=5$ by adopting 2 ZTF filters, our choice of $\eta=1$ results in a $P(\rho) \propto (1-\rho^2)^{3/2}$, which mildly favors small correlation coefficients. 
The vector of standard deviations $\boldsymbol{\sigma}$ is defined as 
\begin{equation}
    \boldsymbol{\sigma} = (\sigma_{t_\mathrm{rise}}, \sigma_{\alpha_g}, \sigma_{\alpha_r}, \sigma_{\ln A_g}, \sigma_{\ln A_r})^\mathsf{T},
\end{equation}
and all these components are drawn from half-Cauchy priors.

This multivariate Gaussian framework enables direct sampling of the population-level mean, scatter, and covariance of all elements in $\boldsymbol{\Theta}$, while the correlation coefficient matrix can be directly computed from the Cholesky decomposition $\mathbf{L}$ as
\begin{equation}
    \mathbf{R} = \mathbf{L}\,\mathbf{L}^\mathsf{T},
\end{equation}
where each element $\mathrm{R}_{ij}$ represents the correlation coefficient between the $i$-th and $j$-th parameters in $\boldsymbol{\theta}_k$.
This can be extended to any linear combination of these parameters, such as $\alpha_g - \alpha_r$, which can be used to quantify the early-time color evolution of SNe\,Ia: a positive (negative) $\alpha_g - \alpha_r$ value indicates a faster (slower) rise in the bluer $g$ band compared to the redder $r$ bands, corresponding to a red$\rightarrow$blue (blue$\rightarrow$red) color evolution at early times. 
Under this framework, the population mean and scatter of $\alpha_g - \alpha_r$ are given by
\begin{equation}
    \mu_{\alpha_g - \alpha_r} = \mu_{\alpha_g} - \mu_{\alpha_r},
\end{equation}
\begin{equation}
    \sigma_{\alpha_g - \alpha_r}^2 = \sigma_{\alpha_g}^2 + \sigma_{\alpha_r}^2 - 2\,\mathrm{Cov}(\alpha_g, \alpha_r),
\end{equation}
while its correlation with other parameters (such as $t_\mathrm{rise}$) is
\begin{equation}\label{eq:rho_t_alpha_g-r}
    \rho(t_\mathrm{rise}, \alpha_g - \alpha_r) = \frac{\mathrm{Cov}(t_\mathrm{rise}, \alpha_g) - \mathrm{Cov}(t_\mathrm{rise}, \alpha_r)}{\sigma_{t_\mathrm{rise}}\sigma_{\alpha_g - \alpha_r}}.
\end{equation}
As such, we can also directly draw samples from the posteriors of $\mu_{\alpha_g - \alpha_r}$, $\sigma_{\alpha_g - \alpha_r}$, and $\rho(t_\mathrm{rise}, \alpha_g - \alpha_r)$.

In Figure~\ref{fig:hierarchical_model}, we present a graphical representation of our hierarchical Bayesian model. The prior distributions of all parameters are summarized in Table~\ref{tab:priors}.

\begin{deluxetable}{lcll}
\tablenum{1}
\tablecaption{Prior Distributions of Model Parameters \label{tab:priors}}
\tablehead{
\colhead{Parameter} & \colhead{Unit} & \colhead{Description} & \colhead{Prior}
}
\startdata
\multicolumn{4}{c}{\textbf{General Parameters}} \\
$C_{k, b}$ & flux unit\tablenotemark{a} & Baseline flux in filter $b$ for SN $k$ & Uniform($-50, 50$) \\
[1ex]
\multicolumn{4}{c}{\textbf{Hierarchical Bayesian Model}} \\
$\mu_{t_\mathrm{rise}}$ & day & Mean rise time of the population & Uniform($5, 35$)\tablenotemark{b} \\
$\mu_{\alpha_b}$ & \nodata & Mean power-law index of the population in filter $b$ & Uniform($1, 4$) \\
$\mu_{\ln A_b}$ & \nodata & Mean log-amplitude of the population in filter $b$ & Uniform($0, \ln(10^{3})$) \\
$\sigma_{t_\mathrm{rise}}$ & day & Scatter in rise time of the population & HalfCauchy($1.5$)\tablenotemark{c} \\
$\sigma_{\alpha_b}$ & \nodata & Scatter in power-law index of the population in filter $b$ & HalfCauchy($0.3$) \\
$\sigma_{\ln A_b}$ & \nodata & Scatter in log-amplitude  of the population in filter $b$ & HalfCauchy($0.5$) \\
$\mathbf{L}$ & \nodata & Cholesky decomposition of correlation matrix & LKJ(1) \\
[1ex]
\multicolumn{4}{c}{\textbf{Unpooled Model}} \\
$t_{\mathrm{rise}, k}$ & day & Rise time of SN $k$ & Uniform($0, 40$) \\
\multirow{2}{*}{$\alpha_{k, b}$} & 
\multirow{2}{*}{\nodata} & 
\multirow{2}{*}{Power-law index of SN $k$ in filter $b$} & 
Uniform(1, 5) \textbf{(Uniform)}\\  
 & & & Exp(1) + 1 \textbf{(Max Entropy)}\\
$\ln A_{k, b}$ & \nodata & Log-amplitude of SN $k$ in filter $b$ & Uniform($0, \ln(10^3)$) \\
\enddata
\tablenotetext{a}{The flux has been normalized such that 100 flux units correspond to the brightness of a $2\times10^{28}\,\mathrm{erg\,s^{-1}\,Hz^{-1}}$ source with a flat SED at the assigned redshift, observed in the respective filter.}
\tablenotetext{b}{The uniform priors for $\mu_{t_\mathrm{rise}}$ and $\mu_{\alpha_b}$ are slightly narrower than those for individual events ($t_\mathrm{rise}$ and $\alpha_b$). This prevents population-level parameters from taking extreme values unsupported by the data in the sampling process to improve the convergence of the NUTS sampler.}
\tablenotetext{c}{The Half-Cauchy distribution allows for a wide range of scatter values (compared to, e.g., a Half-Normal distribution) while still favoring smaller values.}
\end{deluxetable}

\subsection{Likelihood and the Volume-projection Bias} \label{sec:setup_likelihood}

\begin{figure*}
    \centering
    \includegraphics[width=0.83\linewidth]{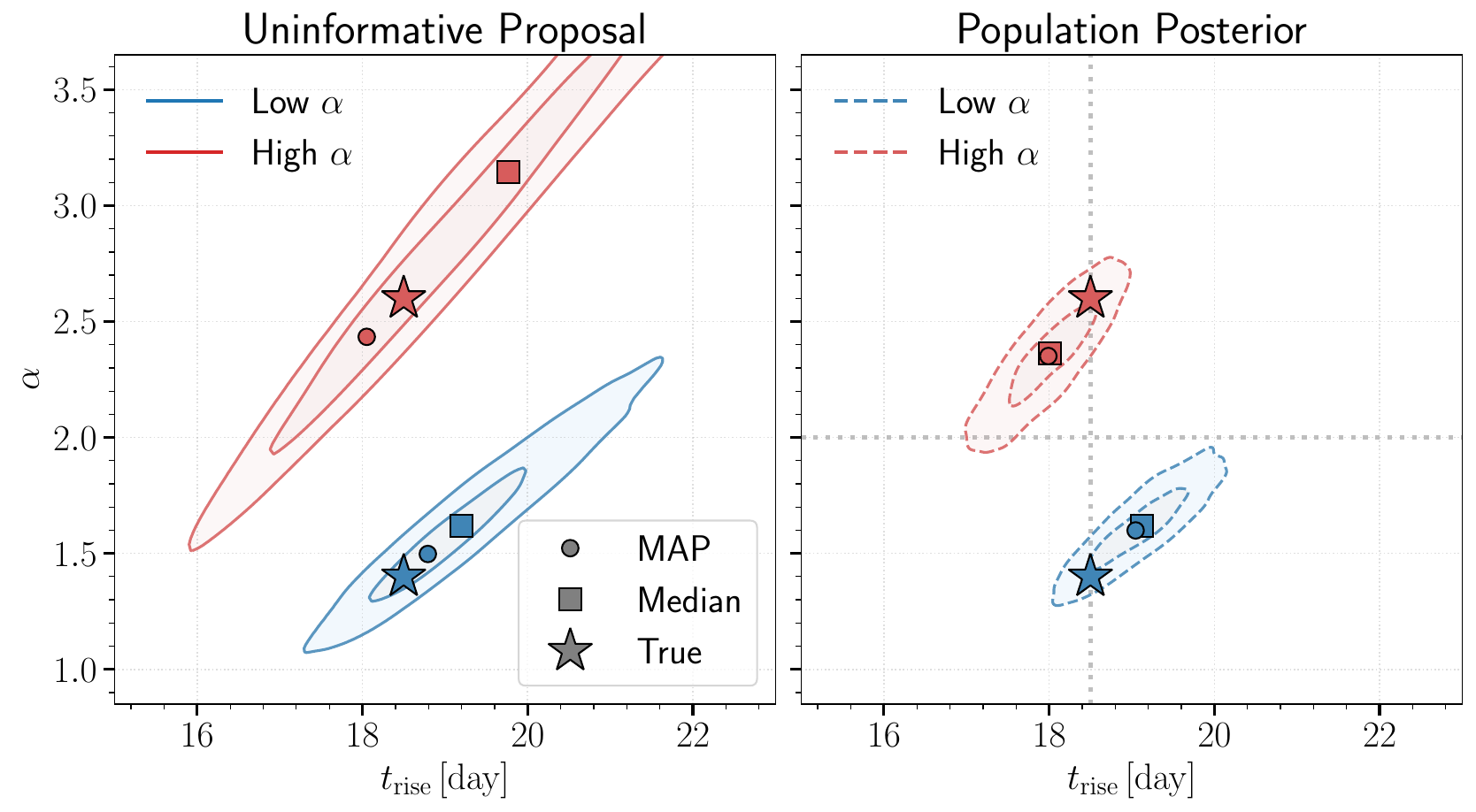}
    \caption{Demonstration of how the highly asymmetric likelihood surface of the power-law model impacts the $t_\mathrm{rise}$ and $\alpha$ posteriors for two synthetic SNe ($\alpha=2.6$, red; $\alpha=1.4$, blue), alongside the effect of Bayesian shrinkage.
    \textit{Left:} Joint posteriors sampled with uninformative uniform priors. The MAP (circles) and marginalized median (squares) estimates are compared to the true parameters (stars). The high-$\alpha$ SN shows an extended posterior tail, creating a severe separation between the MAP and marginalized median. \textit{Right:} Joint posteriors resampled using Gaussian population priors. The posteriors become significantly more compact and symmetric, drawing the marginalized medians closer to both the MAP estimates and the ground truth, but also introducing a bias towards the population mean (dotted lines).
    }
    \label{fig:bayesian_shrinkage}
\end{figure*}

Assuming individual photometric measurements are independent and normally distributed, the log-likelihood for the entire dataset is
\begin{equation}
    \ln \mathcal{L} = -\frac{1}{2}\sum_{i}\left[
    \frac{(f_{k, b}(t_i) - f_{\mathrm{obs}, i})^2}{\sigma_{\mathrm{obs}, i}^2} + \ln(2\pi\sigma_{\mathrm{obs}, i}^2)
    \right],
\end{equation}
where the summation is over all observations $i(k, b)$ for SN $k$; $f_{k, b}(t_i)$ is the model flux at time $t_i$, and $f_{\mathrm{obs},i}$ is the corresponding observed flux. 

Under the power-law formulation, the likelihood surface is notoriously asymmetric, with $t_\mathrm{rise}$, $\alpha$, and $\ln A$ being highly entangled (e.g., \citealp{alderingRiseTimesHigh2000,gonzalez-gaitanRISETIMENORMAL2011,firthRisingLightCurves2015,zhengEmpiricalFittingMethod2017a}; \citetalias{millerZTFEarlyObservations2020}). This strong degeneracy creates an extended tail toward earlier first-light epochs and steeper power laws. When adopting an uninformative prior, this extended tail occupies a considerably larger volume in the parameter space than the narrow peak near the true parameters. Consequently, marginalizing over the posterior inherently introduces a strong volume-projection bias.

Furthermore, as we demonstrate here, this bias disproportionately affects inferences for high-$\alpha$ SNe. 
We illustrate this using two synthetic light curves generated with $\alpha=2.6$ and $\alpha=1.4$, respectively, while fixing $t_\mathrm{fl}=18.5$\,days and $A=40$. Both events are sampled every day from $-100$ to $-10$\,days in phase (with a random jitter drawn from a uniform $-0.1$ to $0.1$\,days distribution) and a Gaussian noise of $\sigma_\mathrm{obs}=1$. 

When sampled using uninformative uniform priors, the joint ($t_\mathrm{rise}, \alpha$) posteriors for both events exhibit the expected ``banana-shaped'' degeneracy (Figure~\ref{fig:bayesian_shrinkage}, left panel). 
Notably, the high-$\alpha$ event exhibits a much more extended posterior tail, leading to severe separation between the maximum a posteriori (MAP) estimate (the region of the highest localized density) and the marginalized median (the center of the integrated volume).
This occurs because a larger $\alpha$ leads to weaker emission shortly after first light. Consequently, the earliest epochs of the light curve (phenomenologically the ``dark'' phase) remain buried in the noise, providing much weaker constraints on the onset of the explosion.

A population prior mitigates this volume-projection effect by down-weighting the unconstrained tails. 
To test this Bayesian shrinkage effect, we resample the posteriors of these events applying Gaussian population priors for all three parameters: $t_\mathrm{rise}\,[\mathrm{day}]\sim\mathcal{N}(18.5, 1.5^2)$, $\alpha\sim\mathcal{N}(2.0, 0.3^2)$, and $\ln A \sim \mathcal{N}(\ln 40, 0.2^2)$. By design, both synthetic SNe are 2$\sigma$ outliers in $\alpha$ but typical otherwise.
Under these informative priors, the posteriors become significantly more compact and less skewed, drawing the marginalized medians closer to the MAP estimations (Figure~\ref{fig:bayesian_shrinkage}, right panel). 

The downside of the Bayesian shrinkage is that the marginalized medians are inevitably biased towards the population mean, particularly for the events in the population tails.
In particular, the intrinsic asymmetry of the likelihood surface ensures this shrinkage is not symmetric across the population.
Because the sprawling unconstrained tail of the high-$\alpha$ event overlaps so heavily with the $\mu_a=2.0$ population mean, the population prior exerts a much stronger downward pull on it than the upward pull on the more tightly constrained low-$\alpha$ event.
We quantify this parameter-dependent Eddington bias by simulating 100 noise realizations of the light curves, finding that the mean deviation in $\alpha$ is substantially larger for the high-$\alpha$ SN ($0.25\pm0.08$) than for the low-$\alpha$ SN ($0.15\pm0.11$). As we will show in Section~\ref{sec:mock_pwl}, this asymmetric constraint means that even a full hierarchical Bayesian model will systematically underestimate the population mean of $\alpha$ by a small amount, as the model naturally anchors to the better-sampled, shallower risers under the power-law rise formulation.

\subsection{Model Implementation} \label{sec:setup_implementation}

The standard approach to hierarchical Bayesian modeling in astrophysics typically relies on a two-step process. First, individual events are evaluated independently using uninformative default priors, generating a discrete set of \textit{proposal} posterior samples for each event. Second, importance sampling reweights these samples, replacing the uninformative default prior with the hyperparameter-conditioned population prior. The reweighted samples then approximate the overarching \textit{target} posterior distribution of the hyperparameters.

While computationally efficient, especially for expensive likelihoods, this two-step framework is susceptible to biased inference. 
If the \textit{proposal} posteriors are poorly constrained or lack sufficient overlap with the \textit{target} distribution, the reweighting process can introduce severe Monte Carlo noise. Given that $t_\mathrm{rise}$ and $\alpha$ are highly degenerate and feature an asymmetric likelihood surface (Section~\ref{sec:setup_likelihood}), fitting individual SNe\,Ia with uninformative priors can yield highly biased individual posteriors. Consequently, the discrete \textit{proposal} samples may drastically under-represent the true parameter space, ultimately skewing population-level inferences during importance sampling.

However, because our likelihood function is computationally cheap to evaluate, we can bypass this two-step approximation entirely. By leveraging the high sampling efficiency of the No-U-Turn Sampler \citep[NUTS;][]{hoffmanNoUTurnSamplerAdaptively2014}, we instead sample directly from the full posterior distribution, jointly fitting the individual parameters and population-level hyperparameters for the entire dataset simultaneously.

We implement the model in \texttt{NumPyro} \citep{binghamPyro2019, phanComposableEffectsFlexible2019} for the NUTS sampling.
By default, we use 3,000 warm-up steps, increasing this number if the sampled chains fail to converge. 
To optimize the efficiency of the NUTS sampler, we implement a two-stage warm-up strategy. During the first 10\% of the warm-up steps, we apply a simulated annealing phase with gradual likelihood tempering, allowing the sampler to broadly explore the parameter space. The subsequent NUTS chains are then initialized using the parameter medians derived from the final 10--20\% of these simulated annealing iterations. These chains then undergo a standard NUTS warm-up to learn the optimal step size and mass matrix for efficient sampling.
Finally, we evaluate convergence using the Gelman-Rubin statistic ($\hat{R}$) and the effective sample size ($N_\mathrm{eff}$). We require $\hat{R} \le 1.01$ and $N_\mathrm{eff} \ge 10^3$ across all random variables to ensure convergence.

\section{Validation on Synthetic Light Curves} \label{sec:mock_pwl}

\begin{deluxetable}{lcll}
\tablenum{2}
\tablecaption{Population-level Distributions for Mock Light Curve Samples\label{tab:mock_sample}}
\tablewidth{0pt}
\tablehead{
    \colhead{Parameter} & 
    \colhead{Unit} & 
    \colhead{Description} & 
    \colhead{Distribution}
}
\startdata
\multicolumn{4}{c}{\textbf{General Parameters}} \\
$t_\mathrm{rise}$ & day & Rise time & $\mathrm{Normal}(18.5, 1.5^2)$ \\
$\alpha$          & \nodata & Power-law index & $\mathrm{Normal}(2.0, 0.3^2)$ \\
[1ex]
\multicolumn{4}{c}{\textbf{Power-law}} \\
$\ln A$          & \nodata & Log-amplitude & $\mathrm{Normal}(\ln40, 0.2^2)$ \\
$\rho(t_\mathrm{rise}, \alpha)$ & \nodata & Correlation coefficient: $t_\mathrm{rise}$, $\alpha$ & 0 \textit{or} $0.3$ \\
$\rho(t_\mathrm{rise}, \ln A)$ & \nodata & Correlation coefficient: $t_\mathrm{rise}$, $\ln A$ & 0 \textit{or} $-0.3$ \\
$\rho(\alpha, \ln A)$ & \nodata & Correlation coefficient: $\alpha$, $\ln A$ & 0 \\
[1ex]
\multicolumn{4}{c}{\textbf{Curved Power-law}} \\
$\alpha_0$          & \nodata & Initial power-law index & $\mathrm{Normal}(2.0, 0.3^2)$ \\
$\dot \alpha$     & day$^{-1}$ & Curvature parameter & Equation~(\ref{eq:curved_risetime}) \\
[1ex]
\multicolumn{4}{c}{\textbf{Broken Power-law}} \\
$t_\mathrm{rise}/t_b$             & \nodata & Ratio of the rise time and break time & $\mathrm{Uniform}(1.0, 1.5)$ \\
$s$               & \nodata & Smoothing parameter & $\mathrm{Uniform}(0.5, 1.5)$ \\
$\alpha_0$          & \nodata & Initial power-law index & $\mathrm{Normal}(2.0, 0.3^2)$ \\
$\alpha_1$ & \nodata & Final power-law index & Equation~(\ref{eq:broken_risetime}) \\
[1ex]
\multicolumn{4}{c}{\textbf{Power-Law $+$ Flux Excess}} \\
$t_\mathrm{rise}$ & day & Rise time & $18.5$ \\
$\alpha$          & \nodata & Power-law index & $2.0$ \\
$\ln A$          & \nodata & Log-amplitude & $\ln 40$ \\
$A_\mathrm{ex}$      & \nodata & Peak flux of excess & $\mathrm{Uniform}(2, 5)$ \\
$t_\mathrm{ex, fwhm}$ & day    & Duration (FWHM)     & $\mathrm{Uniform}(1, 7)$ \\
\enddata
\end{deluxetable}

In this section, we validate our hierarchical Bayesian model using a suite of mock light curves to demonstrate its ability to recover both individual light-curve parameters and population-level parameters under various conditions, and systematically revisit the impact of fitting individual SNe independently.

We create datasets of mock SNe whose luminosities follow the power-law rise model (Equation~\ref{eq:powerlaw}). For simplicity, we assume a flat spectral energy distribution, so the flux density is identical in both filters at any given time. The light-curve parameters are randomly sampled following the rules listed in Table~\ref{tab:mock_sample}. 
The population means of $t_\mathrm{rise}$ and $\alpha$ are chosen to be representative of normal SNe\,Ia \citepalias[e.g.,][]{millerZTFEarlyObservations2020}, while the selected population mean of $A$ ensures that the mock SNe reach 40\% of the peak flux at $t_\mathrm{pivot}$ on average. To demonstrate the model's ability to recover correlations between parameters, we test two cases for each correlation coefficient listed in Table~\ref{tab:mock_sample}: (i) no correlation; and (ii) moderate correlation between $t_\mathrm{rise}$ and $\alpha$ ($\rho=0.3$), and anti-correlation between $t_\mathrm{rise}$ and $\ln A$ ($\rho=-0.3$).

To construct a realistic mock dataset, we generate synthetic ZTF-like observations for a sample of SNe drawn from a volumetric distribution.
The redshifts are drawn from a uniform spatial distribution up to $z=0.06$, mimicking the volume-complete sample in ZTF DR2 \citep{rigaultZTFSNIa2025}. 
We define a normalized flux scale to anchor the synthetic light curves. Since a power-law rise lacks an intrinsic maximum, we set 100 flux units to correspond to the observed flux of a source with a reference luminosity per unit frequency of $2\times10^{28}\,\mathrm{erg\,s^{-1}\,Hz^{-1}}$ (corresponding to an absolute AB magnitude of $\simeq$$-19$\,mag in the $B$ band) at the randomly drawn redshift in a specific filter. Consequently, the modeled flux and measurement uncertainties are scaled relative to this reference.
Following these, we generate synthetic ZTF observations in $g$ and $r$ bands for 1,000 mock SNe using \texttt{REDBACK} \citep{sarinREDBACKBayesianInference2024}, which samples observation epochs and the corresponding sky noises from the actual ZTF observing history between 2018 and 2020.

\subsection{Impact of Early-Time Data Coverage} \label{sec:mock_coverage}

\begin{figure*}
    \centering
    \includegraphics[width=0.9\linewidth]{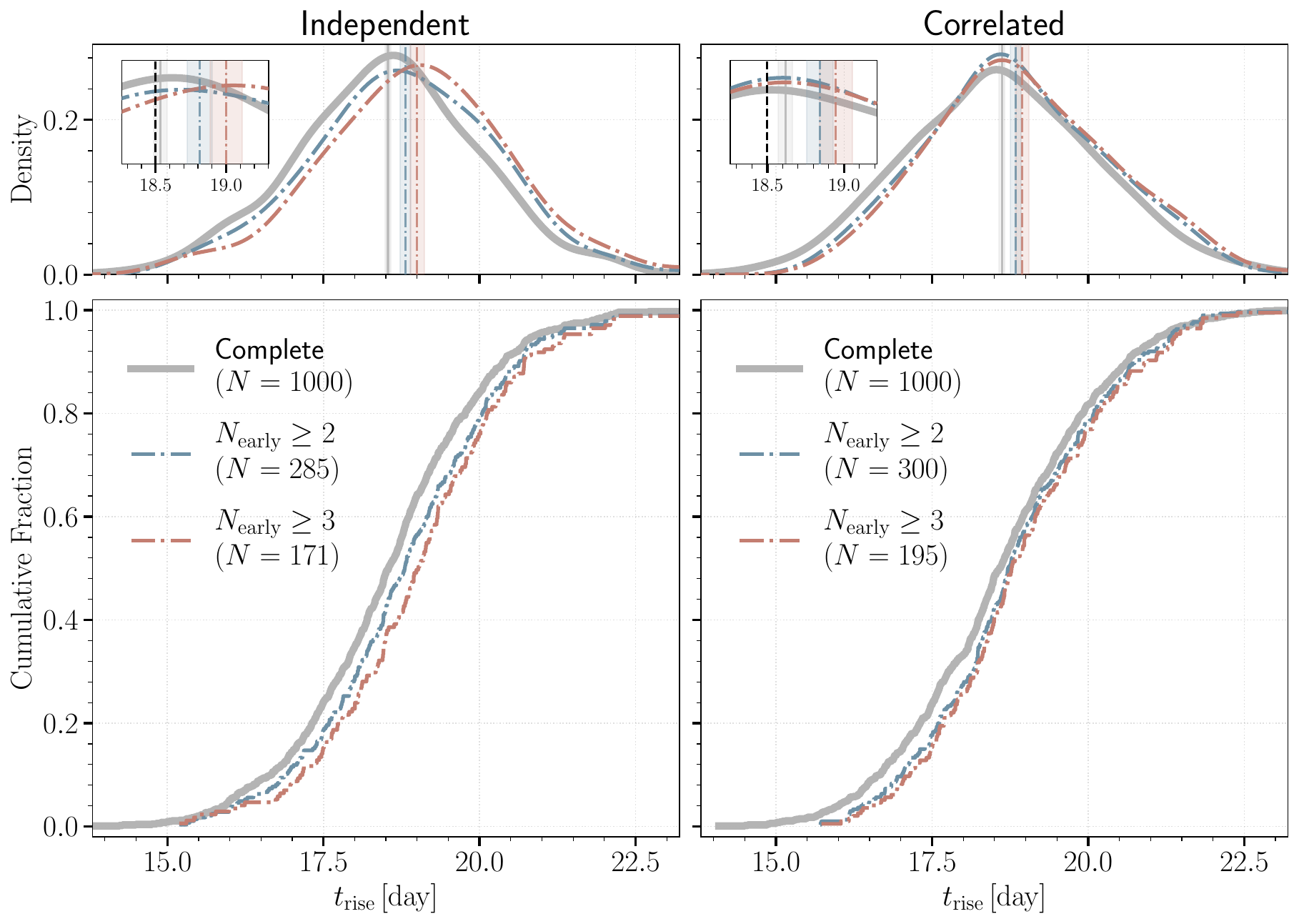}
    \caption{Applying early-time coverage cuts to the full sample biases the distribution of rise times. \textit{Top:} Kernel density estimates (KDEs) of $t_\mathrm{rise}$ in two mock SN\,Ia populations: one where light-curve parameters are drawn independently (\textit{left}) and one where they follow empirical correlations (\textit{right}). The complete simulated sample with minimal early coverage requirements (grey solid line) is compared against subsets surviving strict early-time photometric constraints: $N_\mathrm{early}\ge 2$ (blue) and $N_\mathrm{early}\ge3$ (red). The inset axes zoom in on the central densities, illustrating the systematic shift of the sample means (vertical colored lines) away from the ground-truth population mean ($\mu_{t_\mathrm{rise}}=18.5$\,days; black dashed line) as coverage requirements become more stringent. Shaded regions denote the standard error of the sample mean, derived from the central limit theorem. Bottom panels: Corresponding empirical cumulative distribution functions (ECDFs).}
    \label{fig:quality_cuts}
\end{figure*}

\begin{figure*}[ht!]
    \centering
    \includegraphics[width=\linewidth]{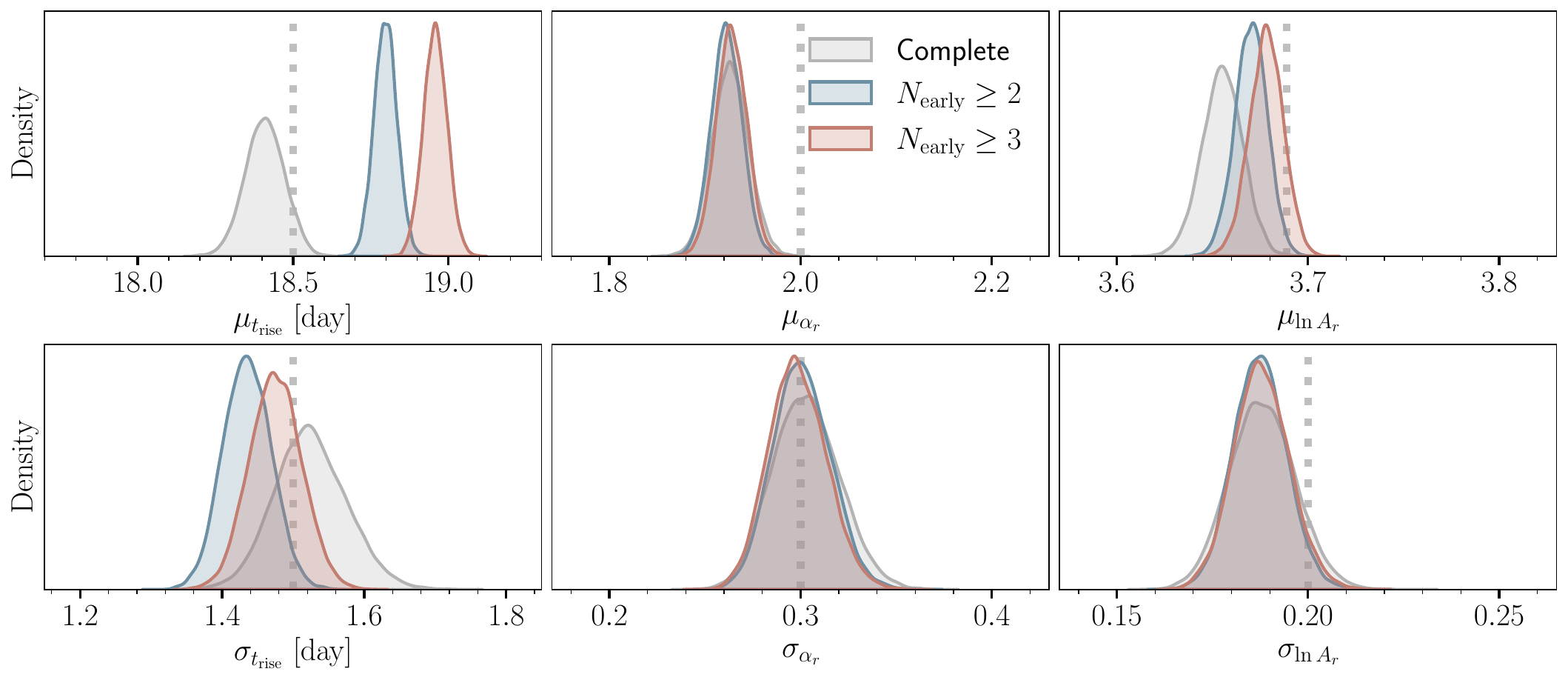}
    \caption{Inferred population-level mean and scatter for $t_\mathrm{rise}$, $\alpha_r$, and $\ln A_r$ across three mock samples with varying early-time coverage requirements. The inferred parameter distributions are compared against the true input values (dashed lines). The results remain consistent across the three samples, except for the mean rise time, $\mu_{t_\mathrm{rise}}$, which is biased toward longer durations when stricter early-time coverage requirements are applied, and the mean rise index, $\mu_{\alpha_r}$, which is systematically underestimated.}
    \label{fig:mock_hbm_qual_mean_scatter}
\end{figure*}

Our primary dataset has 1,000 mock SNe with an extremely loose coverage requirement:
\begin{itemize}
    \item $\ge$2 nights\footnote{In the sample selection process, we stack all synthetic observations within each night to improve the S/N. During the light-curve modeling we still fit all data points.} with signal-to-noise ratio (S/N) $>5$ detections in either filter prior to the peak;\footnote{The ``peak epoch'' for the power-law rise model serves merely as a reference point, defined such that $t_\mathrm{fl}$ occurs $t_\mathrm{rise}$ days prior.}
    \item Detections in both the $g$ and $r$ bands.
\end{itemize}
A significant fraction of these SNe have few detections within the first few days after the first light, in which case we can validate the capability in the hierarchical Bayesian model to constrain the population-level parameters, by properly weighting the contribution of each SN according to its light-curve coverage and S/N.

To mimic the construction of a ``golden'' subset with decent early-time coverage, in which the rise morphology of an individual SN can be well constrained, we apply the following selection criteria:
\begin{itemize}
    \item $\ge$$N_\mathrm{early}$ ($=$$2, 3$) nights with $\mathrm{S/N}>5$ detections before $-10$\,days relative to the peak in both filters;
    \item $\ge$10 nights with pre-explosion non-detections between $-100$\,days and $- 25$\,days in both filters.
\end{itemize}
Here we test two values of $N_\mathrm{early}$. Even for $N_\mathrm{early}=3$, the early-time coverage requirement is still significantly less stringent than the selection in \citetalias{millerZTFEarlyObservations2020}, which only included SNe\,Ia in ZTF high-cadence fields, such that there are typically $\gtrsim$5 detections in both filters before $-10$\,days relative to the peak. The second criterion ensures that the baseline flux offset $C$ can be well constrained in both passbands.

In Figure~\ref{fig:quality_cuts}, we present the rise-time distributions for the full sample alongside subsets with varying early-time coverage requirements. Applying even mild early-time quality cuts significantly skews the rise-time distribution. For power-law light curves with uncorrelated parameters, the median rise time artificially increases from the true population mean of 18.5\,days to $\sim$18.8\,days and $\sim$19.0\,days for $N_\mathrm{early}=2$ and $3$, respectively.
This selection bias occurs because SNe with shorter rise times inherently yield fewer observations prior to $-10$\,days, making them disproportionately likely to fail coverage criteria.
When the light-curve parameters are intrinsically correlated, the selection effect on the $t_\mathrm{rise}$ distribution manifests slightly differently but remains substantial. Notably, given a moderate positive correlation between $t_\mathrm{rise}$ and $\alpha$, the sample mean of even the complete mock dataset is slightly biased toward longer rise times. This excludes the true population mean at $>$$2\sigma$ confidence, although the absolute bias remains small ($\sim$0.1\,days).

In Figure~\ref{fig:mock_hbm_qual_mean_scatter}, we compare the population-level mean and scatter of $t_\mathrm{rise}$, $\alpha_r$, and $\ln A_r$ in the three mock samples with different early-time coverage requirements, as inferred by the hierarchical Bayesian model. Following \citetalias{millerZTFEarlyObservations2020}, we set the truncation threshold to be 40\% unless otherwise specified. We will revisit the impact of this choice in Section~\ref{sec:mock_frac} when fitting a power-law model to the early-time light curves with other functional forms.
The selection bias in $t_\mathrm{rise}$ discussed above is reflected in the inferred population mean: $\mu_{t_\mathrm{rise}}$ is systematically biased toward longer rise times as the early-time coverage requirement becomes more stringent, even though other inferred parameters are consistent with each other across the three samples.
We also identify a systematic bias in $\mu_{\alpha_r}$, which is underestimated by $\sim$0.07, consistent with the asymmetric Bayesian shrinkage discussed in Section~\ref{sec:setup_likelihood}. With ZTF-like cadence and S/N, restricting the analysis to better-sampled SNe does not mitigate this bias. Given the strong degeneracy between the rise index and the amplitude, $\mu_{\ln A_r}$ is also underestimated by $\sim$0.04 across all three samples.
Nonetheless, the inferred population scatters for all three parameters are consistent with the true input values, demonstrating the model's power to constrain the diversity of the population.

\subsection{Independent versus Hierarchical Fits} \label{sec:mock_indep}
\begin{figure*}[ht!]
    \centering
    \includegraphics[width=\linewidth]{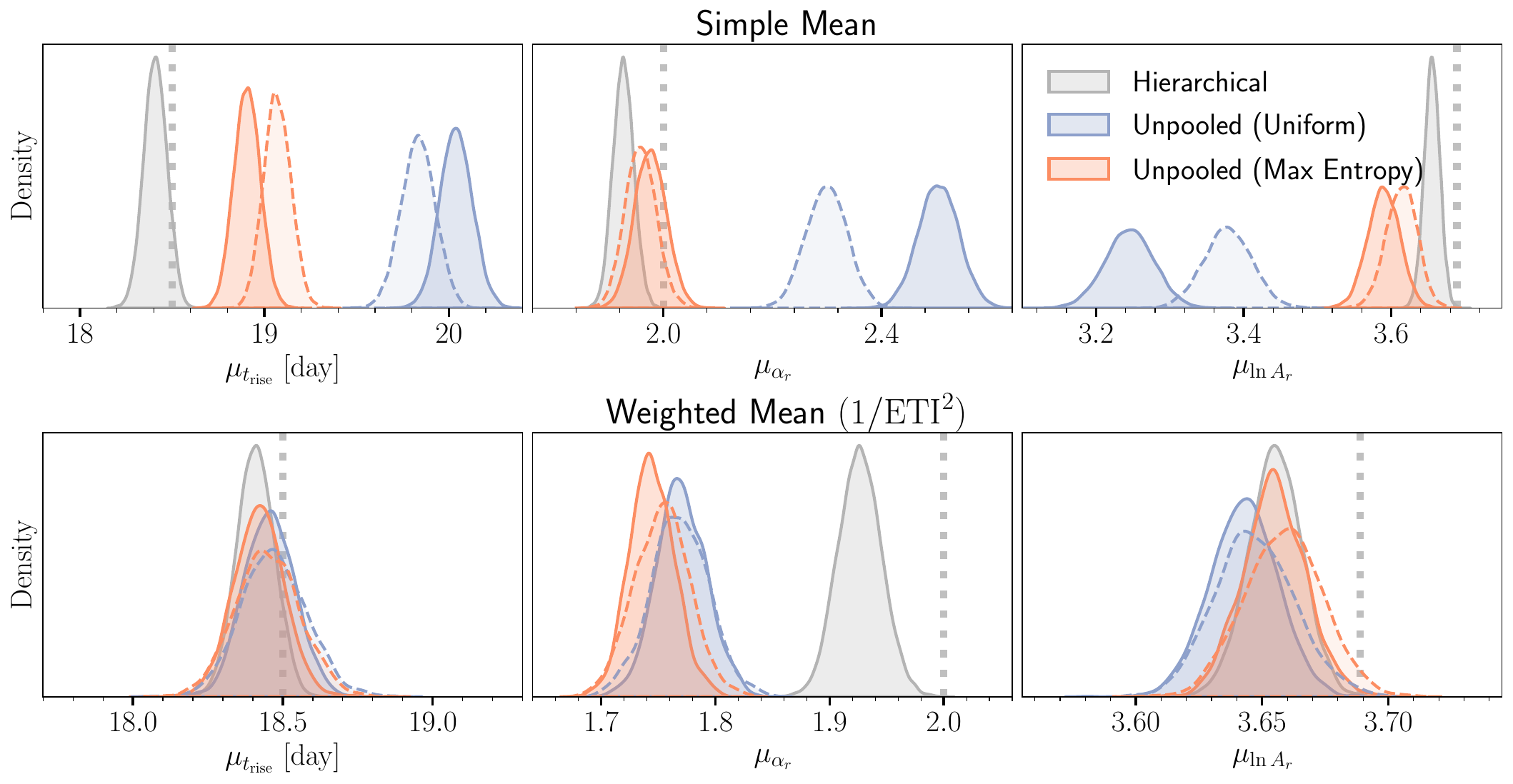}
    \caption{Aggregating the outcomes of individual fits (the unpooled method; colored profiles), either by naively stacking the individual posterior samples (\textit{top panels}) or by reweighting each SN by the inverse variance of its parameters (\textit{bottom panels}), generally yields biased inferences compared to the hierarchical Bayesian model (gray profiles). 
    These biases are sensitive to both prior choices (blue and orange profiles, corresponding to different priors on $\alpha$) and sample selection (solid and dashed profiles, representing the $N_\mathrm{early}\ge2$ and $N_\mathrm{early}\ge3$ subsets, respectively, alongside good baseline coverage).
        }
    \label{fig:mock_hbm_unpooled_mean}
\end{figure*}

\begin{figure*}[ht!]
    \centering
    \includegraphics[width=\linewidth]{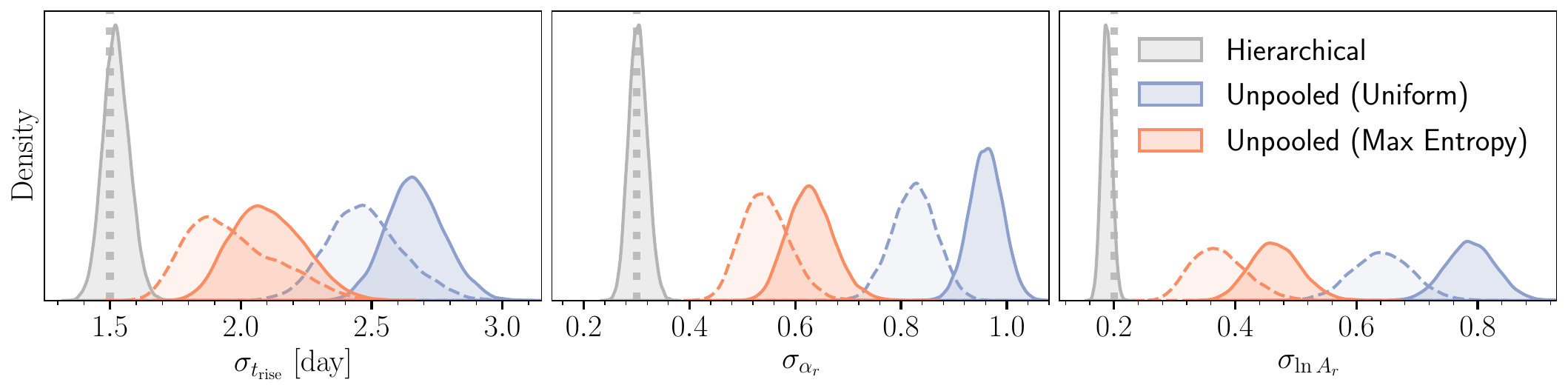}
    \caption{Hierarchical Bayesian model recovers the population-level scatter in synthetic data. In contrast, naively stacking the population properties from individual fits (the unpooled method) generally overestimates the intrinsic scatter. The formatting matches Figure~\ref{fig:mock_hbm_unpooled_mean}.}
    \label{fig:mock_hbm_unpooled_scatter}
\end{figure*}

\begin{figure*}[ht!]
    \centering
    \includegraphics[width=\linewidth]{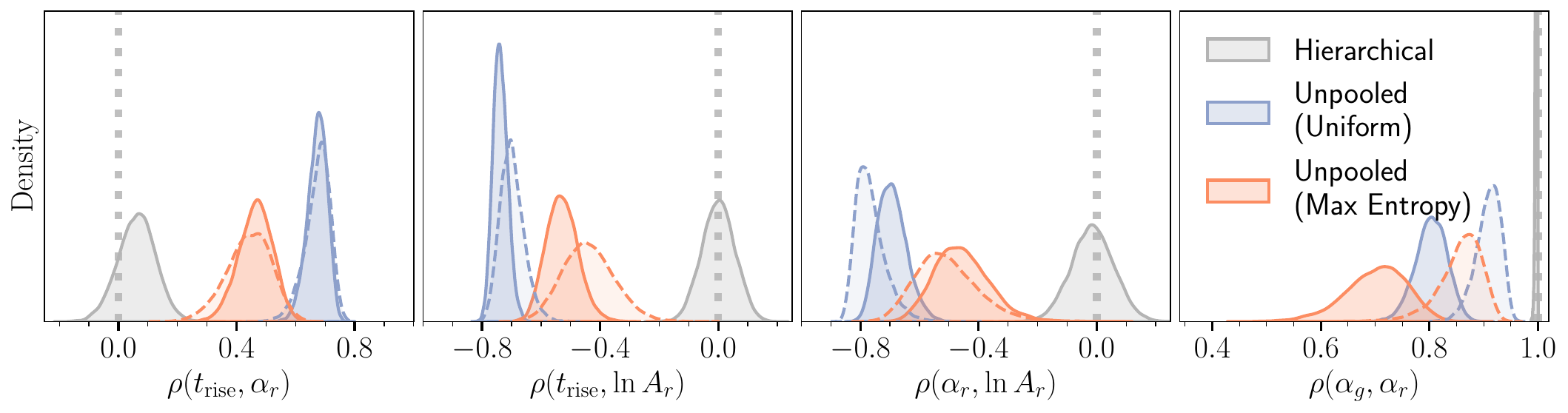}
    \caption{Hierarchical Bayesian modeling recovers the true parameter correlations in synthetic data. The formatting matches Figure~\ref{fig:mock_hbm_unpooled_mean}. The hierarchical model correctly identifies the lack of correlation between $t_\mathrm{rise}$ and $\alpha$, and the perfect correlation between $\alpha_g$ and $\alpha_r$. In contrast, aggregating individual fits yields spurious, moderately strong correlations between $t_\mathrm{rise}$ and $\alpha$, and severely underestimates the correlation between the band-specific power-law indices.}
    \label{fig:mock_hbm_unpooled_corr}
\end{figure*}

\begin{figure*}[ht!]
    \centering
    \includegraphics[width=\linewidth]{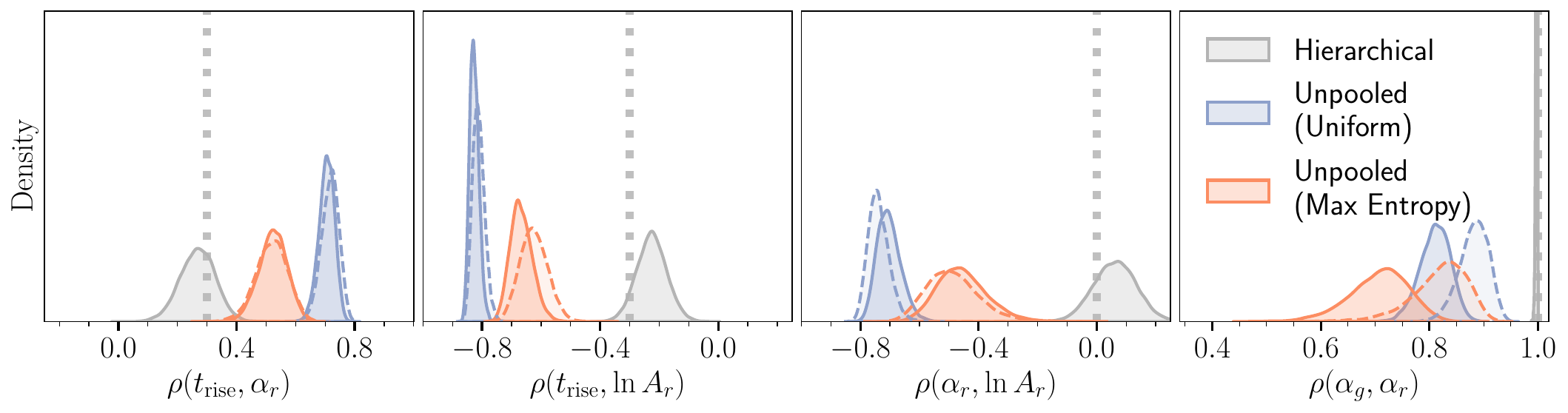}
    \caption{Same as Figure~\ref{fig:mock_hbm_unpooled_corr}, but displaying the case where the mock data are generated with a moderate correlation between $t_\mathrm{rise}$ and $\alpha$ ($\rho=0.3$), and a moderate anti-correlation between $t_\mathrm{rise}$ and $\ln A$ ($\rho=-0.3$). Hierarchical Bayesian modeling successfully recovers the amplitudes of true correlations.}
    \label{fig:mock_corr_hbm_unpooled_corr}
\end{figure*}

In comparison to the hierarchical model, we also build a series of unpooled models assuming no underlying population structure, effectively reducing the graphical model in Figure~\ref{fig:hierarchical_model} to the lower two layers. They represent the classic approach of fitting individual SNe independently and then aggregating the population properties from these individual fits.

The priors for this independent modeling are also listed in Table~\ref{tab:priors}.
We test two prior specifications for the power-law rise index $\alpha$: (i) a uniform prior; and (ii) a shifted exponential distribution, $1+\mathrm{Exp(1)}$. The latter represents the maximum entropy prior for $\alpha>1$ with an expectation value of $E[\alpha]=2$. 
Because sparsely sampled light curves render meaningless individual parameter estimates, we limit this test to the SN subsets with baseline coverage and $N_\mathrm{early}\ge2$ as defined in Section~\ref{sec:mock_coverage}. This process also mimics the common practice of performing analysis on a ``golden'' subset with better early-time coverage and data quality \citep[e.g.,][]{haydenRISEFALLTYPE2010,deckersConstrainingTypeIa2022}. As a result, the SN subsets evaluated by the hierarchical and unpooled models are not identical.

Figure~\ref{fig:mock_hbm_unpooled_mean} compares the population means of $\alpha$, $t_\mathrm{rise}$, and $\ln A$ inferred by the hierarchical Bayesian model (applied to the full sample; gray profiles) against those from the unpooled models (for the $N_\mathrm{early}\ge2$ and $N_\mathrm{early}\ge3$ subsets; solid and dashed color-shaded profiles, respectively). 

For the unpooled models, we first approximate the posterior distribution of the population mean by directly averaging the individual SN posterior samples. This approach, displayed in the upper panels of Figure~\ref{fig:mock_hbm_unpooled_mean}, effectively assigns equal weight to all SNe regardless of data quality and coverage. 

To downweight the contribution of poorly constrained SNe, previous studies have often adopted a weighted mean approach. For instance, \citetalias{millerZTFEarlyObservations2020} weighted each SN by the inverse square of its 68\% equal-tailed credible interval (ETI; the difference between the 84th and 16th percentiles). To evaluate this approach, we perform a bootstrap experiment. For each parameter, we first compute the per-object ETI width from its marginal posterior to define a normalized selection weight $w_k \propto 1 / \mathrm{ETI}_k^2$. We then construct a bootstrap distribution of the weighted mean over 2,000 iterations. In each iteration, $N_\mathrm{SN}$ objects are drawn with replacement according to these weights; a single posterior draw is taken from each, and their simple mean is recorded. The resulting posterior distribution of this weighted mean is displayed in the lower panels of Figure~\ref{fig:mock_hbm_unpooled_mean}.

The unpooled models with the population means aggregated as simple means of individual posterior samples yield the most biased estimates, which are also highly sensitive to the prior choices. The uniform prior on $\alpha$ leads to a significant overestimation of $\mu_\alpha$, which subsequently inflates $\mu_{t_\mathrm{rise}}$ due to their intrinsic degeneracy. This artificial inflation is directly driven by the variance from the unconstrained tails. The maximum entropy prior produces less biased means because its expectation roughly matches the true population. Its overestimation of $\mu_{t_\mathrm{rise}}$ is primarily driven by the selection bias due to the early-time coverage cuts. Increasing the early-time coverage requirement from $N_\mathrm{early}\ge2$ to $N_\mathrm{early}\ge3$ only partially mitigates the bias in the population mean and scatter, while the selection bias in $t_\mathrm{rise}$ becomes more severe.

Adopting a weighted mean makes the estimates less sensitive to prior choices and reduces, though does not eliminate, the biases. Consequently, both $\mu_{t_\mathrm{rise}}$ and $\mu_\alpha$ are underestimated. Interestingly, the inferred $\mu_{t_\mathrm{rise}}$ appears consistent with the true value of 18.5\,days only because the statistical inference bias and the sample selection bias serendipitously cancel each other out. As demonstrated in Section~\ref{sec:setup_likelihood}, the posterior of a high-$\alpha$ SN is inherently more extended than that of a low-$\alpha$ SN under uninformative priors (Figure~\ref{fig:bayesian_shrinkage}), even with identical observation cadence and S/N. The weighting scheme therefore disproportionately downweights high-$\alpha$ SNe, systematically underestimating $\mu_\alpha$. While the hierarchical model experiences a similar bias, its regularizing population prior significantly reduces the weight disparity between high-$\alpha$ and low-$\alpha$ events.

Furthermore, the weighted mean approach is limited to estimating the population mean and provides no straightforward method to constrain the population scatter or correlation structure.
We therefore use the standard deviation of the aggregated SN draws to construct a posterior of the population scatter.
Crucially, this procedure does not deblend the intrinsic population scatter from the measurement uncertainties of individual SNe. In contrast, the hierarchical model explicitly separates these components by parameterizing a population-level dispersion term independently from the likelihood. Figure~\ref{fig:mock_hbm_unpooled_scatter} compares the population-scatter posterior inferred by the hierarchical framework against the sample-scatter posterior derived from the unpooled approach. As expected, the unpooled method yields systematically broader distributions across all parameters because it reflects total dispersion rather than intrinsic scatter alone.

The same conflation corrupts the inferred correlation structure. Figure~\ref{fig:mock_hbm_unpooled_corr} compares the population-correlation posteriors from both models using mock data intentionally generated with zero intrinsic parameter covariances. For the unpooled fits, we compute the sample correlation coefficient $\rho$ across the aggregated SN draws, yielding a full posterior distribution. 
The hierarchical Bayesian model accurately recovers the ground truth, finding no evidence of correlation between $t_\mathrm{rise}$, $\alpha$, and $\ln A$, and correctly identifies the perfect correlation between the $g$- and $r$-band power-law indices. 
In contrast, the unpooled fits yield moderately strong, artificial correlations between $t_\mathrm{rise}$, $\alpha$, and $\ln A$. As discussed in Section~\ref{sec:setup_likelihood}, this bias arises from the inherent degeneracy of the power-law model. Because individual SN posteriors elongate along this degeneracy, aggregating these extended estimates creates the illusion of a population-level trend. Limiting the unpooled fits to better-sampled SNe does not significantly mitigate this bias, as the parameter degeneracy is intrinsic to the power-law formulation itself.
Additionally, the unpooled approach significantly underestimates the correlation between $\alpha_g$ and $\alpha_r$. In practice, an SN is not necessarily sampled equally in both filters, under which circumstances the unpooled method is unaware of the intrinsic correlation between $\alpha_g$ and $\alpha_r$, leading to an underestimation of their correlation coefficient. This is confirmed by the fact that when limiting the unpooled fits to the subset of SNe with $N_\mathrm{early}\ge3$ (which typically have better coverage in both filters), the inferred correlation between $\alpha_g$ and $\alpha_r$ is improved, albeit still underestimated compared to the hierarchical model.

We also test the case where the light-curve parameters are intrinsically correlated. The hierarchical approach still successfully recovers the true correlation coefficients, while the unpooled method again yields biased estimates, overestimating both the correlation between $t_\mathrm{rise}$ and $\alpha$ ($\rho=0.3$) and the anti-correlation between $t_\mathrm{rise}$ and $\ln A$ ($\rho=-0.3$), as shown in Figure~\ref{fig:mock_corr_hbm_unpooled_corr}.

In summary, aggregating individual fits from a sample with heterogeneous data quality could lead to biased estimates of population-level properties. These biases are highly sensitive to the prior choices, and as demonstrated, parameter degeneracies in individual fits can strongly distort the inferred parameter correlation structure. These findings underscore the importance of adopting a hierarchical Bayesian framework to robustly infer the physical properties of early-time SN light curves.

\section{Impact of Complex Light-Curve Morphologies} \label{sec:mock_misspecification}
Although a simple power-law effectively describes the initial rise of many SNe\,Ia, the flux growth rate gradually declines as the SNe approach peak luminosity. 
Additionally, some events display prominent early-time flux excesses that diverge entirely from a smooth rise. In this section, we examine the behavior and robustness of our parameterization when applied to light curves featuring these complex morphological deviations.

\subsection{Truncation Thresholds} \label{sec:mock_frac}
The definition of the ``early-time'' phase, during which the power-law approximation is considered valid, is often ambiguous. Typically, the light-curve data are truncated based on either a fixed number of days prior to maximum light \citep[e.g.,][]{riessRiseTimeNearby1999}, or a specific fraction of the peak flux \citepalias[e.g.,][]{millerZTFEarlyObservations2020}.

Alternative parametric forms have been used to characterize the deviation from a single power-law rise, which enables fitting the entire rise of the light curve without relying on ad hoc truncation. \citet{firthRisingLightCurves2015} proposed a curved power-law model by adding a linear correction term to the exponent \citep[see also][]{vallelyHighcadenceEarlytimeObservations2021,fausnaughFourYearsType2023}, i.e.,
\begin{equation}\label{eq:curved_powerlaw}
    f(t)\propto (t-t_\mathrm{fl})^{\alpha_0(1+\dot \alpha(t-t_\mathrm{fl}))},
\end{equation}
implying that the assumptions made in deriving the simple power-law rise weaken as the SN evolves. 
Alternatively, \citet{zhengVERYYOUNGTYPE2013} introduced a broken power-law model, in which the light curve transitions from a power-law rise to a power-law decline at a break time $t_b$:
\begin{equation}\label{eq:broken_powerlaw}
    f(t) \propto \left(\frac{t-t_\mathrm{fl}}{t_b}\right)^{\alpha_0} \left[1+\left(\frac{t-t_\mathrm{fl}}{t_b}\right)^{s\alpha_1}\right]^{-1/s},
\end{equation}
where $s$ is a smoothing parameter that controls the sharpness of the transition. This formulation is motivated by the broken power-law velocity evolution \citep[characterized by $\alpha_1$ and $\alpha_2$; see][]{zhengEmpiricalFittingMethod2017} observed in other explosive transients, such as gamma-ray burst afterglows.

Both alternatives characterize the gradual deviation from a simple power-law rise as the effective exponent decreases over time. As such, fitting early-time light curves to a power-law rise effectively yields a mean $\alpha$ over the time range, which is smaller than the original $\alpha_0$. Including later epochs would further pull $\alpha$ down and delay the inferred $t_\mathrm{fl}$, leading to an underestimation of $t_\mathrm{rise}$. 

Here we systematically investigate how the choice of flux truncation thresholds impacts the inferred $t_\mathrm{rise}$, under the assumption that the underlying light curves follow either the curved power-law or broken power-law model.
Following the procedure in Section~\ref{sec:mock_pwl}, we generate two datasets of 1000 mock SN\,Ia light curves, synthesized using either the curved power-law (Equation~\ref{eq:curved_powerlaw}) or broken power-law (Equation~\ref{eq:broken_powerlaw}) models. For both models, $t_\mathrm{rise}$ and $\alpha_0$ are drawn from a multivariate Gaussian distribution, as specified in Table~\ref{tab:mock_sample}. The nuisance parameters in the broken power-law model ($s$ and $t_\mathrm{b}$) are drawn from uniform distributions within reasonable ranges following \citet{zhengEmpiricalFittingMethod2018}, also specified in Table~\ref{tab:mock_sample}.
The remaining parameters in both models ($\dot \alpha$ and $\alpha_1$) can be derived analytically. For the curved power-law, the rise time is
\begin{equation}\label{eq:curved_risetime}
    t_\mathrm{rise}\ [\mathrm{day}] = e^{-1}W\left(e^{-e/\dot\alpha}\right),
\end{equation}
where $W(x)$ is the Lambert W function, allowing us to solve for $\dot\alpha$ given the desired $t_\mathrm{rise}$. For the broken power-law, the rise time is
\begin{equation}\label{eq:broken_risetime}
    t_\mathrm{rise} = t_b\left(\frac{\alpha_1}{\alpha_1 - \alpha_0}\right)^{1/s\alpha_1},
\end{equation}
so we similarly solve for $\alpha_1$ given $t_\mathrm{rise}$, $\alpha_0$, and $t_b$. The maximum flux is computed for each mock SN to normalize the light curve such that $f(t_\mathrm{rise})=100$ flux units. Mock ZTF-like observations are then generated as described in Section~\ref{sec:mock_pwl}.
Finally, we model the mock light curves truncated at 30\%, 40\%, and 50\% of the peak flux using our hierarchical Bayesian model with a simple power-law rise.

\begin{figure}[ht!]
    \centering
    \includegraphics[width=\linewidth]{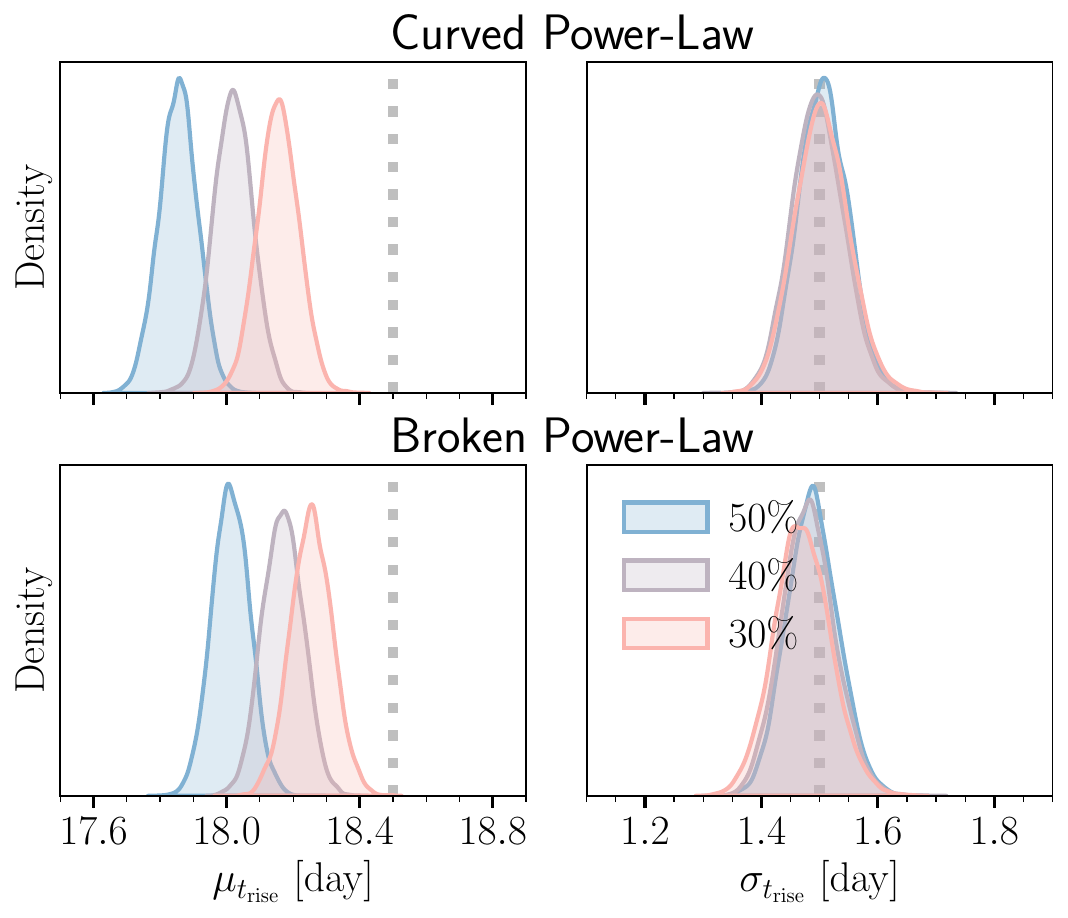}
    \caption{
    Impact of truncating the light curves at different fractions of peak flux on the inferred population parameters for two mock datasets, where the mock light curves are generated with: \textit{Top:} the curved power-law model; \textit{Bottom:} the broken power-law model. For each, we show the inferred posterior distributions of the mean and scatter of $t_\mathrm{rise}$ as a function of the truncation threshold (50\%, 40\%, 30\% of peak flux). The true population mean and scatter of the rise time $t_\mathrm{rise}$ and the initial power-law rise index $\alpha_0$ are indicated by dashed lines. The $t_\mathrm{rise}$ is always underestimated due to model misspecification, but the bias can be mitigated by fitting only data from earlier epochs.
    }
    \label{fig:mock_curved_broken_mean_scatter}
\end{figure}

Figure~\ref{fig:mock_curved_broken_mean_scatter} shows the posterior distributions of the population-level mean and scatter of $t_\mathrm{rise}$ as a function of the truncation threshold for both mock datasets. As expected, the model misspecification leads to a biased $\mu_{t_\mathrm{rise}}$ estimate: when the underlying model is the curved power-law (or broken power-law) and a 50\% truncation threshold, the $\mu_{t_\mathrm{rise}}$ is underestimated by $\sim$0.7\,days ($\sim$0.5\,days), specifically. 
By lowering the truncation thresholds, the biases can be significantly mitigated: with a 30\% truncation threshold, the underestimation of $\mu_{t_\mathrm{rise}}$ is reduced by $\sim$0.3\,days for both underlying models. 
For a robust inference on $\mu_{t_\mathrm{rise}}$, we recommend performing a convergence test by varying this truncation threshold, and limit the fit to earlier epochs if the inferred $\mu_{t_\mathrm{rise}}$ shows a significant dependence on the truncation threshold.
Nevertheless, the scatter in $t_\mathrm{rise}$ is accurately recovered in both cases, as the model can still capture the relative differences in the light-curve shapes across the population, even if the absolute values of $t_\mathrm{rise}$ are biased.

\subsection{Early Flux Excesses} \label{sec:mock_excess}

\begin{figure}
    \centering
    \includegraphics[width=\linewidth]{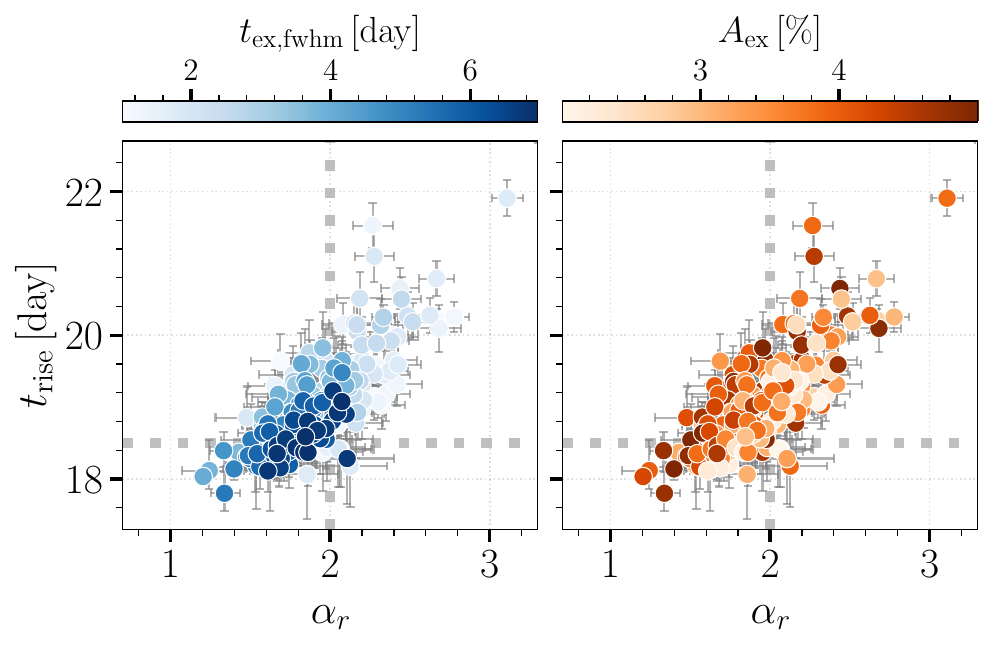}
    \caption{Fitting a single power-law to SNe\,Ia with an early flux excess biases the inferred $t_\mathrm{rise}$ and $\alpha$. Both panels display the inferred parameters for mock light curves generated with an underlying power-law and a Gaussian early flux excess. The mock SNe are color-coded by the excess duration ($t_\mathrm{ex, fwhm}$; \textit{left}) and amplitude ($A_\mathrm{ex}$; \textit{right}). Dashed lines indicate the true values for the underlying power law.}
    \label{fig:mock_excess}
\end{figure}

Early flux excesses that deviate from a smooth power-law rise are prevalent across various SN\,Ia subtypes, including overluminous 03fg-like events \citep{jiangDiscoveryFastestEarly2021,dimitriadisSN2021znyEarly2023,srivastavLuminousTypeIa2023,siebertGroundbasedJWSTObservations2023} and 91T/99aa-like events \citep{jiangSurfaceRadioactivityInteractions2018,millerEarlyObservationsType2018,deckersConstrainingTypeIa2022,wuCommonOriginNormal2025}, as well as subluminous 02es-like objects \citep{caoStrongUltravioletPulse2015,kromerPeculiarTypeIa2016,millerSpectacularUltravioletFlash2020,burkeBrightUltravioletExcess2021,srivastavUnprecedentedEarlyFlux2023,xiSN2022vqzPeculiar2024}. Although much rarer, early flux excesses have also been captured in normal SNe\,Ia. These deviations signal additional energy sources beyond the outward diffusion of photons from radioactive \Ni decay in the center of the SN ejecta. Potential mechanisms include interaction with a companion star \citep{kasenSEEINGCOLLISIONSUPERNOVA2009,kumarFirstDayType2025a} or confined CSM \citep{piroUSINGDOUBLEPEAKEDSUPERNOVA2015,noebauerTypeIaSupernovae2016,piroEXPLORINGPOTENTIALDIVERSITY2016}, or the presence of \Ni directly in the outermost ejecta layers \citep{piroWHATCANWE2013,mageeDetermining56Ni2020}. Depending on the physical mechanism, the duration of this excess can range from a brief, day-long spike (e.g., from interactions with confined CSM) to a broader, week-long effect (e.g., from extended surface \Ni clumps).

Fitting a simple power-law to light curves with these early emission components artificially biases the inferred $t_\mathrm{rise}$ and $\alpha$. Conversely, recognizing anomalous parameters in standard model fits can serve as a diagnostic tool to identify flux excesses. To quantify this effect, we generate mock light curves featuring an underlying power-law with fixed parameters ($t_\mathrm{rise}=18.5$\,days, $\alpha = 2.0$, and $\ln A=\ln 40$), combined with a Gaussian early flux excess.
This excess is characterized by a peak intensity ($A_\mathrm{ex}$) ranging from 2\% to 5\% of the maximum flux, and a duration (full width at half maximum; $t_\mathrm{ex, fwhm}$) spanning 1 to 7\,days. Excesses below 2\% of the peak flux are typically undetectable with ZTF-like cadence and S/N, while those above 5\% are relatively rare among normal SNe\,Ia \citep{deckersConstrainingTypeIa2022}.
The center of the excess ($t_\mathrm{ex, 0}$) is fixed at twice the Gaussian standard deviation following first light (see Table~\ref{tab:mock_sample} for the complete parameter distributions).
We generate 1,000 synthetic light curves but restrict our analysis to the events with good baseline and early-time coverage ($N_\mathrm{early}\ge3$; 209 objects) as defined in Section~\ref{sec:mock_coverage}, as the flux excesses will be completely missed in poorly sampled SNe. 
These synthetic SNe are distributed across the same volume up to $z=0.06$. 

To mimic how anomaly detection functions within a broader population of normal SNe\,Ia, most of which lack early flux excesses, we fit the simple power-law model to each mock SN independently, rather than applying the hierarchical model to the entire sample. We do, however, regularize the posterior sampling using a population prior, mirroring the basic experiment in Section~\ref{sec:setup_likelihood}. Specifically, we adopt population priors of $t_\mathrm{rise}\,\mathrm{[day]}\sim \mathcal{N}(18.5, 1.5^2)$, $\alpha\sim \mathcal{N}(2.0, 0.3^2)$, and $\ln A\sim \mathcal{N}(\ln 40, 0.2^2)$, while fixing the correlations between the band-specific parameters to be near-perfect: $\rho(\alpha_g, \alpha_r) = \rho(\ln A_g, \ln A_r) = 0.99$.

Figure~\ref{fig:mock_excess} displays the resulting posterior distributions for $t_\mathrm{rise}$ and $\alpha$, color-coded by the duration and amplitude of the excess. Even with the population prior stabilizing the fits, an early flux excess can drastically skew the inferred parameters. The duration of the excess dictates the direction of the bias, while its amplitude controls the magnitude. Brief excesses ($t_\mathrm{ex,fwhm}\lesssim2$\,days) typically cause the model to strongly overestimate both $\alpha$ and $t_\mathrm{rise}$. Conversely, longer excesses ($t_\mathrm{ex,fwhm}\gtrsim4$\,days) effectively broaden the early emission, resulting in underestimates for $\alpha$, occasionally yielding near-linear rises ($\alpha\gtrsim1$). These long-duration excesses, however, do not significantly bias $t_\mathrm{rise}$. For even broader excesses where the peak lies close to the truncation epoch, the inferred $\alpha$ increases again.

The vast majority of normal SNe\,Ia lack strong early excesses. For instance, \citet{deckersConstrainingTypeIa2022} found that only $18\pm11$\% of normal SNe\,Ia exhibit early flux excesses exceeding 2\% of their peak luminosity, which can easily be missed with the ZTF cadence \citep{mageeDetectionEfficiencyType2022}. We therefore expect these anomalies to only marginally impact the population-level mean and scatter of $\alpha$ and $t_\mathrm{rise}$. By establishing the baseline demographics, individual SNe\,Ia with prominent early excesses can be readily identified as outliers in the $t_\mathrm{rise}$--$\alpha$ plane. Specifically, events featuring brief, early excesses will manifest as outliers with anomalously large $t_\mathrm{rise}$ and $\alpha$, whereas those with extended excesses will present remarkably low $\alpha$ values paired with comparable rise times. Such distinct morphological signatures offer critical insights into the diverse progenitor setups and explosion environments of SNe\,Ia. In \citet{PaperII}, we applied this framework to a large sample of ZTF SNe\,Ia and identified a small number of outliers in the parameter space, which we interpret as candidates for hosting early flux excesses.

We caution, however, that quantifying the true detection efficiency of this approach requires more comprehensive simulations \citep[e.g.,][]{mageeDetectionEfficiencyType2022}. Specifically, incorporating realistic light-curve morphologies for both the underlying baseline and the excess component will be critical to fully evaluate the capabilities of this method.

\section{Inference of Individual Events}\label{sec:ind}

\begin{figure*}
    \centering
    \includegraphics[width=0.9\textwidth]{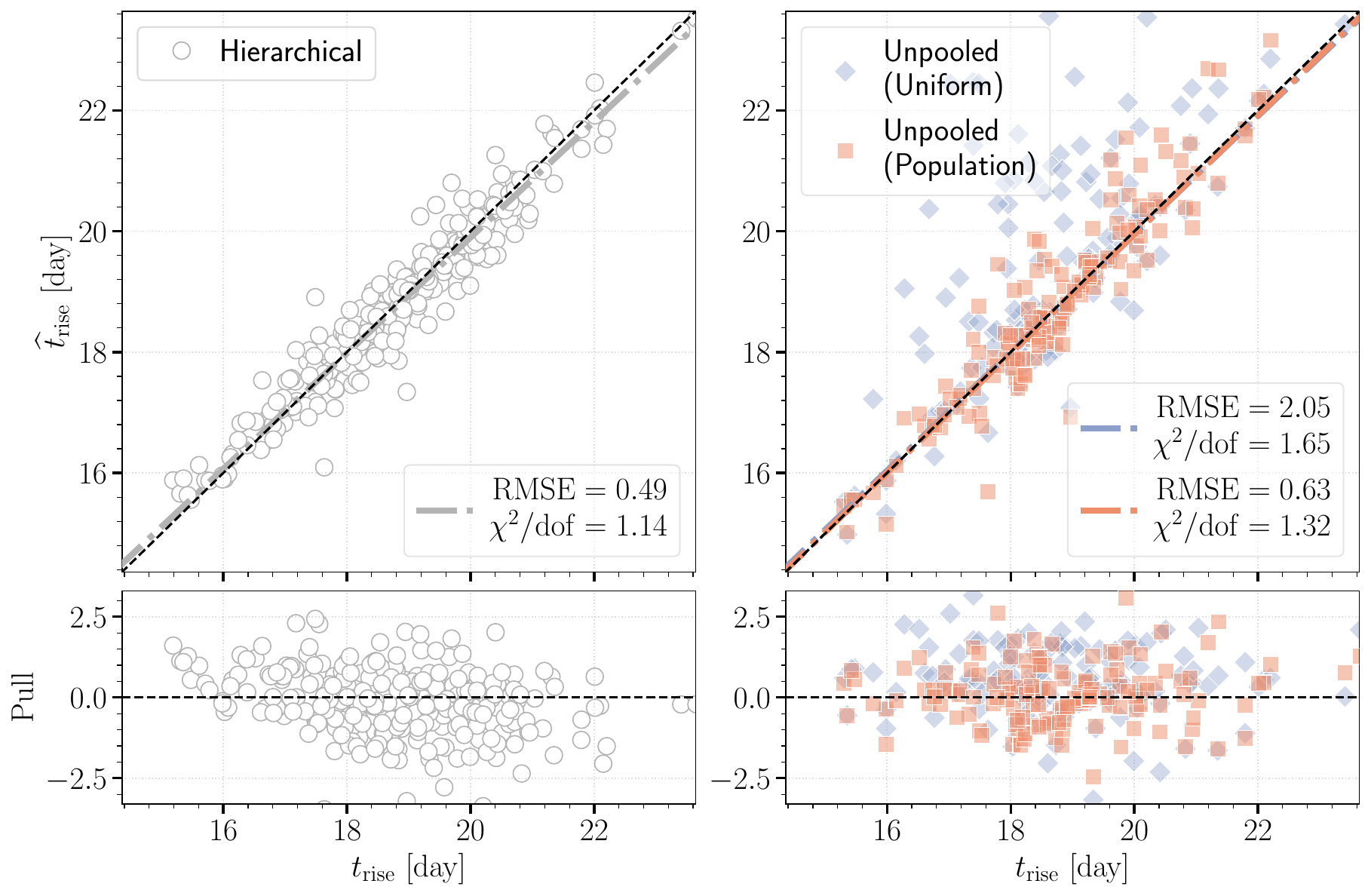}
    \caption{
        Comparison of different modeling approaches in recovering the true rise time of individual SNe\,Ia for the mock sample with uncorrelated parameters and $N_\mathrm{early}\ge2$. \textit{Upper:} Inferred $t_\mathrm{rise}$ (posterior median) versus true values. Error bars are not displayed for clarity. \textit{Lower:} Pulls of the inferred $t_\mathrm{rise}$ (deviations from the true values normalized by uncertainties).
        Dashed black lines indicate the one-to-one relation, and dash-dotted lines represent inverse-variance-weighted linear regressions for each method. The RMSE and reduced $\chi^2$ values are annotated in the upper panels.
        Although the hierarchical model (empty circles) yields low-variance estimates, it exhibits significant shrinkage toward the population mean. Conversely, the unpooled model with uninformative priors (blue) produces largely unbiased estimates, but its scatter is severely inflated by unconstrained tails in the likelihood surface. Applying a population prior that regularizes the full correlation structure alongside the mean and scatter of the nuisance parameter $\ln A$ (orange) achieves a balance between bias and variance. For visual clarity, only half of the events evaluated with the unpooled approaches are displayed.
        }
    \label{fig:mock_individual_t_rise}
\end{figure*}

\begin{figure*}
    \centering
    \includegraphics[width=0.9\textwidth]{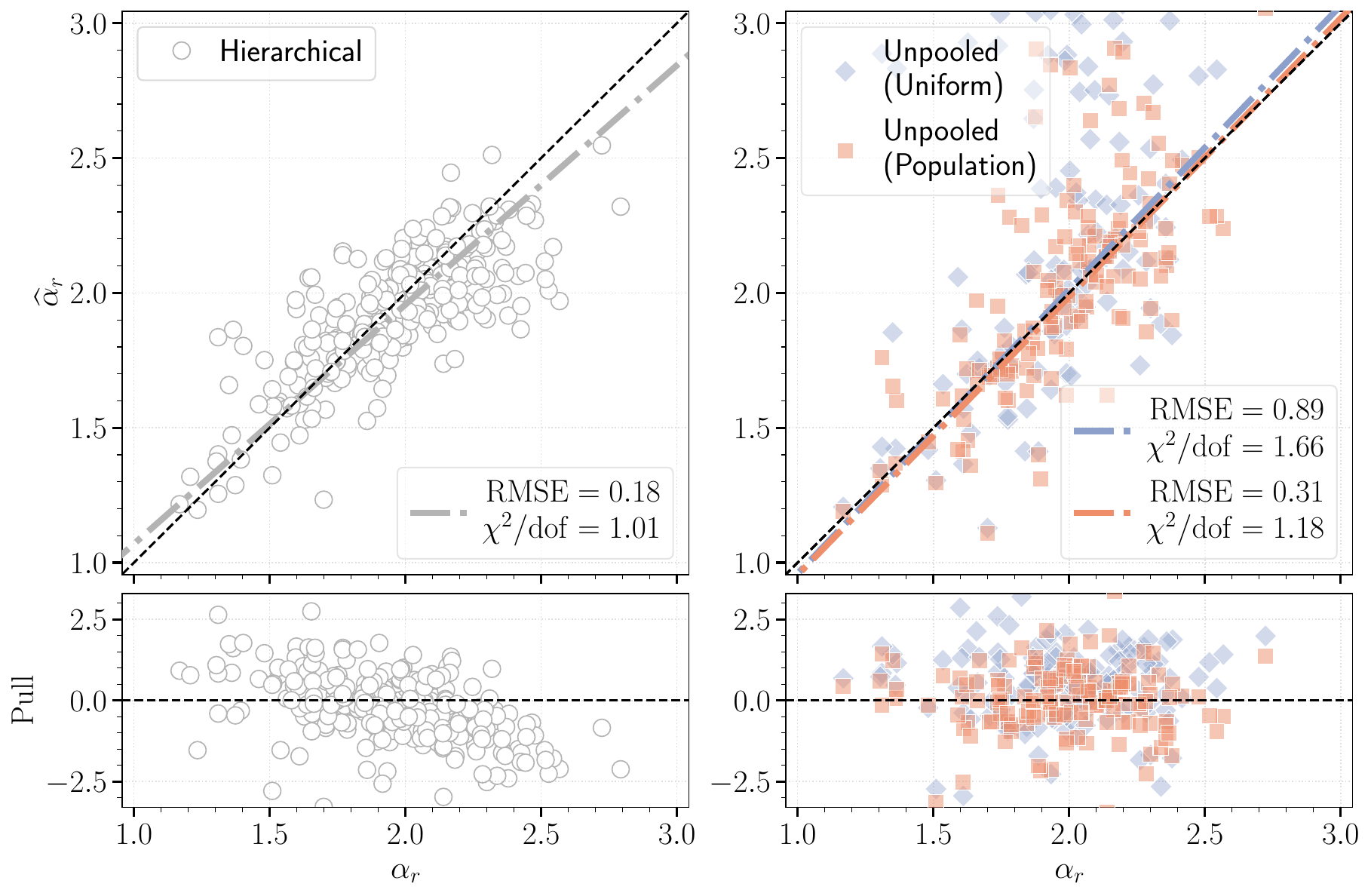}
    \caption{
        Same as Figure~\ref{fig:mock_individual_t_rise}, but showing the inference of the power-law rise index. The custom population prior significantly mitigates the shrinkage effect, while also reducing the scatter compared to the uninformative unpooled model.
    }
    \label{fig:mock_individual_alpha}
\end{figure*}

Although the hierarchical Bayesian framework provides a powerful tool to infer the population-level distribution of early-time light curve parameters, it is not designed to yield precise parameter estimates for individual SNe\,Ia, due to the Bayesian shrinkage effect as we have clearly demonstrated in Section~\ref{sec:setup_likelihood}. In SN\,Ia progenitor studies, however, we are often interested in the early-time properties of a large number of individual events, which can be connected to other observables (e.g., full light-curve parameters, spectral features, host galaxy properties) to identify potential subpopulations and constrain the underlying physical mechanisms. Obtaining low-bias, low-variance estimates for a bulk sample of individual SNe\,Ia with only sub-optimal early-time coverage is thus crucial for these studies.

To evaluate the utility of the hierarchical framework for individual event inference, we compare its ability to recover true SN\,Ia parameters against an unpooled model containing a uniform prior on $\alpha$. This test utilizes the mock sample with uncorrelated parameters and $N_\mathrm{early}\ge2$ as defined in Section~\ref{sec:mock_indep}. Figure~\ref{fig:mock_individual_t_rise} and Figure~\ref{fig:mock_individual_alpha} plot the inferred parameters against their true values, along with inverse-variance-weighted linear regressions to quantify the bias and scatter in the measurements for individual SNe. For visual clarity, we omit individual error bars in the upper panels, opting instead to display the parameter pulls (the deviations from the true values normalized by their uncertainties) in the lower panels.

As expected, the hierarchical model yields low-variance estimates. It achieves a root-mean-square error (RMSE) in both $t_\mathrm{rise}$ and $\alpha$ that is a factor of $\gtrsim$5 smaller than the unpooled model with uninformative priors, alongside a reduced $\chi^2$ ($\chi^2/\mathrm{dof}$, where dof is the number of SNe) close to unity.
However, Bayesian shrinkage toward the population mean is pronounced. This is particularly evident for $\alpha$, where the linear fit slope is roughly 0.9. This effect is less severe for $t_\mathrm{rise}$, but the pulls still exhibit a correlated trend.
Conversely, the unpooled model produces largely unbiased estimations for both $t_\mathrm{rise}$ and $\alpha$, yielding regression slopes near unity. However, this approach leaves many individual events poorly constrained. Driven by the extended tails of the likelihood surface, the variance is heavily inflated. A substantial fraction of events exhibit high estimates for $t_\mathrm{rise}$ and $\alpha$ accompanied by massive uncertainties, rendering them uninformative for single-object studies.

To leverage the population-level information learned from the hierarchical modeling while mitigating the bias from Bayesian shrinkage, we introduce a population prior. This prior retains the full correlation structure learned by the hierarchical model but discards the inferred mean and scatter for the parameters of physical interest, namely $t_\mathrm{rise}$ and $\alpha$; in practice, they are sampled from the same uniform distributions as the unpooled models, yet constrained by the correlation structure. The scatter and mean of the nuisance parameter $\ln A$, however, are retained.
This approach significantly mitigates the shrinkage effect, rendering regression slopes close to unity for both parameters, while also reducing the scatter compared to the uninformative unpooled model, reducing the RMSE for $t_\mathrm{rise}$ and $\alpha$ by a factor of $\sim$3 and $\sim$2, respectively. Correspondingly, the reduced $\chi^2$ values are also significantly improved for both parameters, albeit still larger than unity due to the remaining bias from the population prior.
This improvement is primarily driven by the fact that the population prior on $\ln A$ and the correlation structure effectively regularizes the unconstrained likelihood tails, yet does not directly bias the parameters of interest.

We hence propose a hybrid approach for inferring both the population-level distribution and individual parameters of early-time SN\,Ia light curves: (i) model the entire sample with the hierarchical Bayesian framework to reliably constrain the population-level distribution of $t_\mathrm{rise}$, $\alpha$, and $\ln A$; (ii) fit each SN independently with a population prior on the nuisance parameter $\ln A$ as well as the correlation structure derived from the hierarchical model, which achieves a better balance between bias and variance for individual parameter estimates.

\section{Conclusions}\label{sec:conclusions}
We introduced a hierarchical Bayesian framework to model the early-time light curves of SNe\,Ia.
The classic power-law parameterization of early-time flux, $f(t)\propto (t-t_\mathrm{fl})^\alpha$, features a notoriously asymmetric likelihood surface, which complicates the inference of population-level properties from a sample of individual fits.
By adopting a multivariate Gaussian distribution as a population prior from which individual SNe are drawn, our framework reliably constrains the underlying demographics of early-time light curves while naturally accounting for measurement uncertainties and parameter covariances. 

Through extensive testing on synthetic power-law models, we demonstrated that this hierarchical approach dramatically reduces the severe biases inherent in naively aggregating individual fits. We highlight the following key advantages:
\begin{itemize}
    \item The hierarchical model naturally down-weights sparsely sampled SNe and noisy measurements, enabling high-fidelity inference from large samples with minimal data-quality cuts. This framework mitigates the prevalent selection bias toward longer rise times that plagues strictly filtered ``golden'' subsets. This capability paves the way for population-level analyses on volume-complete, albeit sparsely sampled, datasets from future surveys like Rubin/LSST and Roman.
    \item Regularization from the population prior effectively reduces the volume-projection effects due to the highly asymmetric likelihood surface of the power-law model. This yields much more accurate estimates of the population-level mean and scatter, avoiding the severe biases and prior-sensitivity of naively aggregated individual fits, even though the inherent asymmetry still imparts a minor bias on the inferred rise index.
    \item By directly modeling the population covariance structure, the hierarchical framework accurately recovers true parameter correlations. In contrast, aggregating independent fits falls prey to the intrinsic degeneracies of the power-law parameterization, generating strong but spurious correlations between $t_\mathrm{rise}$, $\alpha$, and $\ln A$. The unpooled approach also severely underestimates cross-band correlations (e.g., between $g$- and $r$-band rise indices) when dealing with unevenly sampled multi-band data.
\end{itemize}

In reality SN light curves are more complex than a simple power-law rise. Using synthetic light curves generated from more sophisticated parametric models to approximate the deviation from a power-law rise, we found that the inferred population-level mean of $t_\mathrm{rise}$ is mildly underestimated due to model misspecification, but the scatter is recovered for both rise models we considered. We recommend performing a convergence test by varying the truncation threshold to ensure a robust inference on $t_\mathrm{rise}$, and limiting the fit to earlier epochs if the inferred $\mu_{t_\mathrm{rise}}$ shows a significant dependence on the truncation threshold.

Should the underlying light curve feature an early flux excess, fitting a simple power-law can significantly bias the inferred $t_\mathrm{rise}$ and $\alpha$. We showed that the direction of the bias is primarily determined by the duration of the excess, while its amplitude controls the magnitude of the bias. 
Given that only a small fraction of normal SNe\,Ia exhibit strong early flux excesses, these anomalies should only marginally impact the population-level mean and scatter of $\alpha$ and $t_\mathrm{rise}$. Instead, individual SNe\,Ia with significant early excesses can be readily identified as outliers in the $t_\mathrm{rise}$--$\alpha$ plane, which can provide critical insights into the diverse progenitor setups and explosion environments of SNe\,Ia.

In the context of individual event inference, a full population prior can lead to significant Bayesian shrinkage toward the population mean, thus biasing the parameter estimates for individual SNe\,Ia, particularly for those in the tails of the population distribution. Nevertheless, by introducing a population prior on the nuisance parameter $\ln A$, which is strongly degenerate with both $t_\mathrm{rise}$ and $\alpha$, we can mitigate this shrinkage effect, achieving a better balance between bias and variance for individual parameter estimates.

Our hierarchical Bayesian framework provides a robust method for inferring the population-level distributions and individual parameters of early-time SN\,Ia light curves. These parameters can then be combined with other observables (e.g., full light-curve parameters, spectral features, and host-galaxy properties) to identify potential subpopulations and map underlying physical mechanisms. In an accompanying paper \citep{PaperII}, we apply this methodology to the volume-complete ZTF SN\,Ia Data Release 2 \citep[DR2;][]{rigaultZTFSNIa2025}. Utilizing a sample of 972 SNe\,Ia, we characterize the demographics of their early-time emission. Furthermore, this framework can be readily extended to other SN subtypes (e.g., SNe\,Ib/c and SNe\,II from massive stars) whose early rises are similarly described by power-law models, albeit with different typical timescales and indices, which will transform our understanding of early-time SN demographics and their connections to progenitor systems across the broader SN landscape. 
The software implementation of our model is publicly available at \url{https://github.com/slowdivePTG/Early_Ia_Rise}.

\software{\texttt{ArviZ} \citep{martinArviZModularFlexible2026}, \texttt{Astropy} \citep{astropycollaborationAstropyCommunityPython2013,astropycollaborationAstropyProjectBuilding2018,astropycollaborationAstropyProjectSustaining2022}, \texttt{JAX} \citep{jax2018github}, \texttt{NumPyro} \citep{binghamPyro2019, phanComposableEffectsFlexible2019}, \texttt{REDBACK} \citep{sarinREDBACKBayesianInference2024}}

\begin{acknowledgments}
We thank Kate Maguire, Tom{\'a}s E. M{\"u}ller-Bravo, and Jesper Sollerman for fruitful discussions on this work.

C.L. and~A.A.M.~are supported by DoE award \#\,DE-SC0025599, while A.A.M.~is also supported by Cottrell Scholar Award \#\,CS-CSA-2025-059 from Research Corporation for Science Advancement.

This work used computing resources provided by Northwestern University and the Center for Interdisciplinary Exploration and Research in Astrophysics (CIERA). This research was supported in part through the computational resources and staff contributions provided for the Quest high performance computing facility at Northwestern University which is jointly supported by the Office of the Provost, the Office for Research, and Northwestern University Information Technology.
\end{acknowledgments}

\bibliography{sn,facility,ML,cross_ref}{}
\bibliographystyle{aasjournalv7}


\end{document}